\documentstyle[12pt]{article}

\def\be{\begin{equation}}
\def\ee{\end{equation}}
\def\LG{LG}

\def\Gra(\S){L}
\def\Ver{{\rm Ver}}
\def\R{{\rm R}}

\def\Ad{{\rm Ad}}
\def\div{{\rm div}}

\def\Cyl{{\rm Cyl}}

\def\0{\emptyset}
\def\C*{$C^\star-{\rm algebra}$}
\def\S{M}

\def\A{{\cal A}}
\def\B{\overline{\A/\G}}
\def\ab{\overline{\A}}
\def\ag{{\A/\G}}
\def\agb{\B}
\def\o{\omega}
\def\Diff{{\rm Diff}}
\def\G{{\cal G}}
\def\H{{\cal H}}
\def\HA{{\cal HA}}
\def\HAbar{\overline{\cal HA}}
\def\HG{{\cal HG}{}}
\def\P{{\cal P}}
\def\D{{\cal D}}
\def\Ga{\Gamma}
\def\X{{\cal X}}

\def\Xb{{\overline \X}}

\def\C{{\cal C}}
\def\P{{\cal P}}

\def\Ab{{\overline \A}}

\def\a{\alpha}  
   
\def\g{\gamma}  

\def\:{\ :}
\def\Gb{{\overline \G}}

\def\xz{x_o}
\def\={\ =\ }

\def\RR{{\rm I\kern -.2em R}}
\def\Comp{{\mathchoice
{\setbox0=\hbox{$\displaystyle\rm C$}\hbox{\hbox to0pt
{\kern0.4\wd0\vrule height0.9\ht0\hss}\box0}}
{\setbox0=\hbox{$\textstyle\rm C$}\hbox{\hbox to0pt
{\kern0.4\wd0\vrule height0.9\ht0\hss}\box0}}
{\setbox0=\hbox{$\scriptstyle\rm C$}\hbox{\hbox to0pt
{\kern0.4\wd0\vrule height0.9\ht0\hss}\box0}}
{\setbox0=\hbox{$\scriptscriptstyle\rm C$}\hbox{\hbox to0pt
{\kern0.4\wd0\vrule height0.9\ht0\hss}\box0}}}}
\def\Co{{\mathchoice
{\setbox0=\hbox{$\displaystyle\rm C$}\hbox{\hbox to0pt
{\kern0.4\wd0\vrule height0.9\ht0\hss}\box0}}
{\setbox0=\hbox{$\textstyle\rm C$}\hbox{\hbox to0pt
{\kern0.4\wd0\vrule height0.9\ht0\hss}\box0}}
{\setbox0=\hbox{$\scriptstyle\rm C$}\hbox{\hbox to0pt
{\kern0.4\wd0\vrule height0.9\ht0\hss}\box0}}
{\setbox0=\hbox{$\scriptscriptstyle\rm C$}\hbox{\hbox to0pt
{\kern0.4\wd0\vrule height0.9\ht0\hss}\box0}}}}
\newtheorem{theorem}{Theorem}
	\newtheorem{proposition}{Proposition}
	\newtheorem{lemma}{Lemma}

\begin{document}
\baselineskip=20pt

\title{Differential geometry on the space of connections via graphs 
and projective limits}

\author{Abhay Ashtekar${}^{1,2}$ and Jerzy Lewandowski${}^{3,4}$}
\maketitle
\centerline{${}^{1}$Center for Gravitational Physics and Geometry,}
\centerline{Physics Department, Penn State, University Park,
PA 16802-6300.}

\centerline{${}^{2}$Isaac Newton Institute for Mathematical Sciences,} 
\centerline{University of Cambridge, Cambridge CB3 OEH.}

\centerline{${}^{3}$Institute of Theoretical Physics,}
\centerline{Warsaw University, ul Hoza 69, 00-681 Warsaw.}

\centerline{${}^{4}$Erwin Schr\"odinger Institute for Mathematical Physics}
\centerline{Pasteurgasse 6/7, A-1090 Vienna.}
\bigskip

{\Large{\bf Abstract}}
\bigskip

In a quantum mechanical treatment of gauge theories (including general
relativity), one is led to consider a certain completion, $\agb$, of
the space $\ag$ of gauge equivalent connections.  This space serves as
the quantum configuration space, or, as the space of all Euclidean
histories over which one must integrate in the quantum theory.  $\agb$
is a very large space and serves as a ``universal home'' for measures
in theories in which the Wilson loop observables are well-defined.  In
this paper, $\agb$ is considered as the projective limit of a
projective family of compact Hausdorff manifolds, labelled by graphs
(which can be regarded as ``floating lattices'' in the physics
terminology).  Using this characterization, differential geometry is
developed through algebraic methods.  In particular, we are able to
introduce the following notions on $\agb$: differential forms,
exterior derivatives, volume forms, vector fields and Lie brackets
between them, divergence of a vector field with respect to a volume
form, Laplacians and associated heat kernels and heat kernel measures.
Thus, although $\agb$ is very large, it is small enough to be
mathematically interesting and physically useful. A key feature of
this approach is that it does not require a background metric.  The
geometrical framework is therefore well-suited for diffeomorphism
invariant theories such as quantum general relativity.

\goodbreak

\section{Introduction}

Theories of connections are playing an increasingly important role in
the current description of all fundamental interactions of Nature
(including gravity \cite{AA1}). They are also of interest from a
purely mathematical viewpoint. In particular, many of the recent
advances in the understanding of the topology of low dimensional
manifolds have come from these theories.

In the standard functional analytic approach, developed in the context
of the Yang-Mills theory, one equips the space of connections with the
structure of a Hilbert-Riemann manifold (see, e.g., \cite{AM}). This
structure is gauge-invariant. However, the construction uses a fixed
Riemannian metric on the underlying space-time manifold.  For
diffeomorphism invariant theories --such as general relativity-- this
is, unfortunately, a serious drawback. A second limitation of this
approach comes from the fact that, so far, it has led to relatively
few examples of interesting, gauge-invariant measures on spaces of
connections, and none that is diffeomorphism invariant. Hence, to deal
with theories such as quantum general relativity, a gauge and
diffeomorphism invariant extension of these standard techniques is
needed.

For the functional integration part of the theory, such an extension
was carried out in a series of papers over the past two years [3-8].
(For earlier work with the same philosophy, see \cite{KK}.)  The
purpose of this article is to develop differential geometry along the
same lines.  Our constructions will be generally motivated by certain
heuristic results in a non-perturbative approach quantum gravity based
on connections, loops and holonomies. \cite{AA3, RS}. Reciprocally,
our results will be useful in making this approach rigorous
\cite{ALMMT2, AA2} in that they provide the well-defined measures and 
differential operators that are needed in a rigorous treatment.  There
is thus a synergetic exchange of ideas and techniques between the
heuristic and rigorous treatments.

As background material, we will first present some physical
considerations and then discuss our approach from a mathematical
perspective.

Fix an n-dimensional manifold $\S$ and consider the space $\A$ of
smooth connections on a given principal bundle $B(\S, G)$ over $\S$.
Following the standard terminology, we will refer to $G$ as the
structure group and denote the space of smooth vertical automorphisms
of $B(\S,G)$ by $\G$. This $\G$ is the group of {\it local} gauge
transformations. If $\S$ is taken to be a Cauchy surface in a
Lorentzian space-time, the quotient $\ag$ serves as the physical
configuration space of the classical gauge theory. If $\S$ represents
the Euclidean space-time, $\ag$ is the space of physically distinct
classical histories. Because of the presence of an infinite number of
degrees of freedom, to go over to quantum field theory, one has to
enlarge $\ag$ appropriately. Unfortunately, since $\ag$ is non-linear,
with complicated topology, a canonical mathematical extension is not
available. For example, the simple idea of substituting the smooth
connections and gauge transformations in $\ag$ by distributional ones
does not work because the space of distributional connections does not
support the action of distributional local gauge transformations.

Recently, one such extension was introduced \cite{AI} using the basic
representation theory of $C^\star$-algebras. The ideas underlying this
approach can be summarized as follows. One first considers the space
$\HA$ of functions on $\ag$ obtained by taking finite complex linear
combinations of finite products of Wilson loop functions $W_{\a}(A)$
around closed loops $\a$.  (Recall that the Wilson loop functions are
traces of holonomies of connections around closed loops; $W_{\a}(A) =
{\rm Tr}\ {\cal P}\ \oint_{\a} A dl$. Since they are gauge invariant,
they project down unambiguously to $\ag$.) $\HA$ can then be completed
in a natural fashion to obtain a $C^\star$-algebra $\HAbar$. This is
the algebra of configuration observables. Hence, to obtain the Hilbert
space of physical states, one has to select of an appropriate
representation of $\HAbar$.  It turns out that every cyclic
representation of $\HAbar$ by operators on a Hilbert space is of a
specific type \cite{AI}: The Hilbert space is simply $L^2(\agb, \mu)$
for some regular, Borel measure $\mu$ on a certain completion $\agb$
of $\ag$ and, as one might expect of configuration operators, the
Wilson loop operators act just by multiplication. Therefore, the space
$\agb$ is a candidate for the required extension of the classical
configuration. To define physically interesting operators, one needs
to develop differential geometry on $\agb$.  For example, the momentum
operators would correspond to suitable vector fields on $\agb$ and the
kinetic term in the Hamiltonian would be given by a Laplacian. The
problem of introducing these operators is coupled to that of finding
suitable measures on $\agb$ because these operators have to be
essentially self-adjoint on the underlying Hilbert space.

>From a mathematical perspective, $\agb$ is just the Gel'fand spectrum
of the Abelian $C^\star$-algebra $\HAbar$; it is a compact, Hausdorff
topological space and, as the notation suggests, $\ag$ is densely
embedded in it.  The basic techniques for exploring the structure of
this space were introduced in \cite{AL1}. It was shown that $\agb$ is
very large: In particular, every connection on {\it every}
$G$-bundle over $\S$ defines a point in $\agb$. (Note incidentally
that this implies that $\agb$ is independent of the initial choice of
the principal bundle $B(\S, G)$ made in the construction of the
holonomy algebra $\HA$.)  Furthermore, there are points which do not
correspond to {\it any} smooth connection; these are the generalized
connections (defined on generalized principal $G$-bundles \cite{Le})
which are relevant only to the quantum theory.  Finally, there is a
precise sense in which this space provides a ``universal home'' for
measures that arise from lattice gauge theories \cite{AMM}. In
specific theories, such as Yang-Mills, the support of the relevant
measures is likely to be significantly smaller. For diffeomorphism
invariant theories, on the other hand, there are indications that it
would be essential to use the whole space. In particular, it is known
that $\agb$ admits large families of measures which are invariant
under the induced action of Diff$(\S)$ \cite{AL1, B1, B2, ALMMT2,AL2}
and therefore likely to feature prominently in non-perturbative
quantum general relativity \cite{ALMMT2, AA2}. Many of these are {\it
faithful} indicating that all of $\agb$ would be relevant to quantum
gravity.
 
Thus, the space $\agb$ is large enough to be useful is a variety of
contexts. Indeed, at first sight, one might be concerned that it is
too large to be physically useful. For example, by construction, it
has the structure {\it only} of a topological space; it is not even a
manifold.  How can one then hope to introduce the basic quantum
operators on $L^2(\agb, \mu)$?  In absence of a well-defined manifold
structure on the quantum configuration space, it may seem impossible
to introduce vector fields on it, let alone the Laplacian or the
operators needed in quantum gravity! Is there a danger that $\agb$ is
so large that it is mathematically uninteresting?

Fortunately, it turns out that, although it is large, $\agb$ is
``controllable''. The key reason is that the $C^\star$-algebra
$\HAbar$ is rather special, being generated by the Wilson loop
observables. As a consequence, its spectrum, $\agb$, can also be
obtained as the projective limit of a projective family of compact,
Hausdorff, {\it analytic manifolds} \cite{AL1, B1, B2, MM,
AL2}. Standard projective constructions therefore enable us to induce
on $\agb$ various notions from differential geometry.  Thus, it
appears that a desired balance is struck: While it is large enough to
serve as a ``universal home'' for measures, $\agb$ is, at the same
time, small enough to be mathematically interesting and physically
useful. This is the main message of this paper.

The material is organized as follows. In section 2, we recall from
\cite{AL2} the essential results from projective techniques. In section
3, we use these results to construct three projective families of
compact, Hausdorff, analytic manifolds, and show that $\agb$ can be
obtained as the projective limit of one of these families. Since the
members of the family are all manifolds, each is equipped with the
standard differential geometric structure. Using projective
techniques, sections 4 and 5 then carry this structure to the
projective limits. Thus, the notions of forms, volume forms, vector
fields and their Lie-derivatives and divergence of vector fields with
respect to volume forms can be defined on $\agb$. The vector fields
which are compatible with the measure (in the sense that their
divergence with respect to the measure is well-defined) lead to
essentially self-adjoint momentum operators in the quantum theory. In
section 6, we turn to Riemannian geometry. Given an additional
structure on the underlying manifold $\S$ --called an edge-metric-- we
define a Laplacian operator on the $C^2$-functions on $\agb$ and
construct the associate heat kernels as well as the heat kernel
measures. In section 7, we point out that $\agb$ admits a natural
(degenerate) contravariant metric and use it to introduce a
Laplace-like operator. Since this construction does not use any
background structure on $\S$, the action of the operator respects
diffeomorphism invariance. It could thus define a natural observable
in diffeomorphism invariant quantum theories. Another example is a
third order differential operator representing the ``volume
observable'' in quantum gravity.  Section 8 puts the analysis of this
paper in the context of the earlier work in the subject.

A striking aspect of this approach to geometry on $\agb$ is that its
general spirit is the same as that of non-commutative geometry and
quantum groups: even though there is no underlying differentiable
manifold, geometrical notions can be developed by exploiting the
properties of the {\it algebra} of functions. On the one hand, the
situation with respect to $\agb$ is simpler because the algebra in
question is Abelian. On the other hand, we are dealing with very
large, infinite dimensional spaces. As indicated above, a primary
motivation for this work comes from the mathematical problems
encountered in a non-perturbative approach to quantum gravity
\cite{AA3} and our results can be used to solve a number of these
problems \cite{ALMMT2, AA2}. However, there some indications that, to
deal satisfactorily with the issue of ``framed loops and graphs'' that
may arise in regularization of certain operators, one may have to
replace the structure group $SL(2, C)$ with its quantum version
$SL(2)_q$. Our algebraic approach is well-suited for an eventual
extension along these lines.

\section{Projective Techniques: general framework}

In this section, we recall from \cite{MM,AL2} some general results on
projective limits which will be used throughout the rest of the paper.

We begin with the notion of a projective family. The first object we
need is a set $L$ of {\it labels}. The only structure $L$ has is the
following: it is a partially ordered, directed set. That is, it is a
set equipped with a relation `$\ge$' such that, for all $\g, \g'$ and
$\g''$ in $L$ we have:
\be \label{2.1a}
\g \ge \g\ ;\quad  
\g \ge \g'\ {\rm and}\  \g'\ge \g \Rightarrow \g =\g'\ ; \quad 
\g\ge\g'\ {\rm and}\ \ \g'\ge\g''\ \Rightarrow \g\ge\g''\ ;
\ee
and, given any $\g', \g'' \in L$, there exists $\g \in L$ such that
\be \label{2.1b}
 \g \ge \g' \quad {\rm and} \quad \g \ge \g''\ . \ee
A {\it projective family} $(\X_\g,p_{\g\g'})_{\g,\g'\in L}$ consists
of sets $\X_\g$ indexed by elements of $L$, together with a family of
surjective {\it projections},
\be p_{\g\g'}\:\ \X_{\g'}\rightarrow \X_{\g},\ee
assigned uniquely to pairs $(\g',\g)$ whenever $\g'\ge\g$ 
such that
\be\label{2.3}
p_{\g\g'}\circ p_{\g'\g''} = p_{\g\g''}.\ee 

A familiar example of a projective family is the following. Fix a
locally convex, topological vector space $V$. Let the label set $L$
consist of finite dimensional subspaces $\g$ of $V^\star$, the
topological dual of $V$. This is obviously a partially ordered and
directed set. Every $\g$ defines a unique sub-space $\tilde{\g}$ of
$V$ via: $\tilde{v} \in \tilde{\g}\ \ {\rm iff}\ \ <v, \tilde{v}> =0
\ \forall v\in \g$. The projective family can now be constructed by 
setting $\X_\g = V/\tilde\g$. Each $\X_\g$ is a finite dimensional
vector space and, for $\g'\ge \g$, $p_{\g \g'}$ are the obvious
projections. Integration theory over infinite dimensional topological
spaces can be developed starting from this projective family
\cite{K,DM}. In this paper, we wish to consider projective 
families which are in a certain sense complementary to this example
and which are tailored to the kinematically non-linear spaces of
interest.

In our case, $\X_\g$ will all be topological, compact, Hausdorff
spaces and the projections $p_{\g\g'}$ will be continuous.  The
resulting pairs $(\X_\g, p_{\g \g'})_{\g,\g' \in L}$ are said to
constitute a {\it compact Hausdorff projective family}.  In the
application of this framework to gauge theories, the labels $\g$ can
be thought of as ``floating'' lattices (i.e., which are not
necessarily rectangular) and the members $\X_{\g}$ of the projective
family, as the spaces of configurations/histories associated with
these lattices.  The continuum theory will be recovered in the
(projective) limit as one considers lattices with increasing number
of loops of arbitrary complexity.

Note that in the projective family there will, in general, be no set
${\Xb}$ which can be regarded as the largest, from which we can
project to any of the $\X_\g$. However, such a set does emerge in an
appropriate limit, which we now define. The {\it projective limit}
$\Xb$ of a projective family $(\X_\g, p_{\g\g'})_{\g,\g'\in L}$ is the
subset of the Cartesian product $\times_{\g\in L}\X_\g$ that
satisfies certain consistency conditions:
\be \label{2.4}
\Xb\ :=\ \{(x_\g)_{\g\in L}\in \times_{\g\in L}\X_\g\  
:\ \g'\ge \g \Rightarrow p_{\g\g'}x_{\g'} = x_\g\}. 
\ee
(In applications to gauge theory, this is the limit that gives us the
continuum theory.)  One can show that $\Xb$, endowed with the topology
that descends from the Cartesian product, {\it is itself a compact,
Hausdorff space}.  Finally, as expected, one can project from the
limit to any member of the family: we have
\be\label{2.5}
 p_{\g}\ : \ \Xb \rightarrow \X_{\g}, \ \  
p_{\g} ((x_{\g'})_{\g'\in L}):= x_{\g}\  .\ee

Next, we introduce certain function spaces. For each $\g$ consider
the space $C^0(\X_\g)$ of the complex valued, continuous functions on
$\X_\g$. In the union 
$$\bigcup_{\g\in L} C^0(\X_{\g})$$ 
let us define the following equivalence relation. Given $f_{\g_i}\in
C^0(\X_{\g_i})$, $i=1,2$, we will say:
\be
\label{2.6}  f_{\g_1} \ \sim\ f_{\g_2}\ \ \ {\rm if}\ \
\ p_{\g_1\g_3}^\star \ f_{\g_1}\ =\ p_{\g_2\g_3}^\star\ f_{\g_2}\ee 
for every $\g_3\ \ge \g_1,\g_2$, where $p^\star_{\g_1\g_3}$ denotes the
pull-back map from the space of functions on $\X_{\g_1}$ to the space
of functions on $\X_{\g_3}$. (Note that to be equivalent, it is in
fact sufficient that the equality (\ref{2.6}) holds {\it just for one}
$\g_3\ \ge \g_1,\g_2$.) 

Using the equivalence relation we can introduce the set of 
{\it cylindrical functions} associated with the projective 
family $(\X_\g,p_{\g\g'})_{\g,\g'\in L}$,
\be\label{2.7}
\Cyl^0(\Xb) \ := \ \big( \bigcup_{\g\in L}C^0(\X_\g)\ \big)
\ /\ \sim.\ee
The quotient just gets rid of a redundancy: pull-backs of functions
from a smaller set to a larger set are now identified with the
functions on the smaller set.  Note that an element of $\Cyl^0(\Xb)$
determines, through the projections (\ref{2.5}), a function on
$\Xb$. Hence, there is a natural embedding
$$
\Cyl^0(\Xb)\ \rightarrow\ C^0(\Xb),
$$
which is dense in the sup-norm. Thus, modulo the completion,
$\Cyl^0(\Xb)$ may be identified with the algebra of continuous
functions on $\Xb$ \cite{AL2}. This fact will motivate, in section 3,
our definition of $C^n$ functions on the projective completion.

Next, let us illustrate how one can introduce interesting structures
on the projective limit. Since each $\X_{\g}$ in our family as well as
the projective limit $\Xb$ is a compact, Hausdorff space, we can use
the standard machinery of measure theory on each of these spaces.  The
natural question is: What is the relation between measures on
$\X_{\g}$ and those on $\Xb$?  To analyze this issue, let us begin
with a definition.  Let us assign to each $\g \in L$ a regular Borel,
probability (i.e., normalized) measure, $\mu_\g$ on $\X_{\g}$.  We
will say that this constitutes a {\it consistent family of measures}
if
\be\label{2.8} (p_{\g\g'})_{\star}\ \mu_{\g'}\ =\ \mu_{\g}\ . \ee
Using this notion, we can now characterize measures on $\Xb$
\cite{AL2}:

\begin{theorem}{\rm :} Let $(\X_\g, p_{\g \g'})_{\g\g' \in 
L}$ be a compact, Hausdorff projective family 
and $\Xb$ be its projective limit;

\smallskip
\noindent (a) Suppose $\mu$ is a regular Borel, probability 
measure on $\Xb$. Then $\mu$ defines a consistent family of regular,
Borel, probability measures, given by:
\be\label{2.22}
\mu_\g\:=\ {p_\g}_\star\mu;\ee

\smallskip
\noindent (b) Suppose $(\mu_\g)_{\g,\g'\in L}$ is a consistent family 
of regular, Borel, probability measures. Then there is a unique
regular, Borel, probability measure $\mu$ on $\Xb$ such that
$({p_\g})_\star\ \mu = \mu_\g$;

\smallskip
\noindent(c) $\mu$ is faithful if  $\mu_\g\ :=\ 
({p_\g})_\star\ \mu$  is faithful for every $\g\in L$. 
\end{theorem}

\noindent This is an illustration of the general strategy we
will follow in sections 4-7 to introduce interesting structures on the
projective limit; they will correspond to families of {\it consistent}
structures on the projective family.

\section{Projective families for spaces of connections.}

We will now apply the general techniques of section 2 to obtain three
projective families, each member of which is a compact, Hausdorff,
analytic manifold.

Fix an n-dimensional, analytic manifold $\S$ and a smooth principal
fiber bundle $B(\S, G)$ with the structure group $G$ which we will
assume to be a compact and connected Lie group. Let $\A$ denote the
space of smooth connections on $B$ and $\G$ the group of smooth
vertical automorphisms of $B$ (i.e., the group of local gauge
transformations).  The projective limits of the three families will
provide us with completions $\Ab$, $\Gb$ and $\agb$ of the spaces
$\A$, $\G$ and $\ag$. As the notation suggests, $\agb$ will turn out
to be naturally isomorphic with the Gel'fand spectrum of the holonomy
algebra of \cite{AI}, mentioned in section 1. 

The label set of all three families will be the same. Section 3.1
introduces this set and sections 3.2--3.4 discuss the three
families and their projective limits. The results of this section
follow from a rather straightforward combination of the results of
\cite{AL1,B1,B2,MM,AL2}.  Therefore, we will generally skip the
detailed proofs and aim at presenting only the final structure which
is used heavily in the subsequent sections.

\subsection{Set of labels.}

The set $L$ of labels will consist of graphs in $\S$. To obtain a
precise characterization of this set, let us begin with some
definitions.

By an unparametrized oriented analytic edge in $\S$ we shall mean
an equivalence class of maps
\be 
e\ :\ [0,1]\ \rightarrow\ \S,
\ee 
where two maps $e$ and $e'$ are considered as equivalent if they
differ only by a reparametrization, or, more precisely if $e'$ can be
written as
\be
e'\ =\ e\circ f,\ \ \ {\rm where} \ \ \ f\ :\ [0,1]\ \rightarrow \
[0,1],
\ee
is an analytic orientation preserving bijection. We will also consider
unoriented edges for which the requirement that $f$ preserve
orientation will be dropped.  The end points of an edge will be
referred to as {\it vertices}. (If the edges are oriented, each $e$
has a well-defined initial and a well-defined final vertex.) A
(oriented) {\it graph} $\g$ in $\S$ is a set of finite, unparametrized
(oriented) analytic edges which have the following properties:
\begin{enumerate}
\item every $e\in\g$ is diffeomorphic with the closed interval $[0,1]$;
\item if $e_1,e_2\in\g$, with $e_1\not= e_2$, the intersection $e_1\cap 
e_2$ is contained in the set of vertices of $e_1,e_2$,
\item every $e\in\g$ is at both sides connected with
another element of $\g$.
\end{enumerate}
\noindent 
(Note that the last condition ensures that each graph is closed.)  The
set of all the graphs in $\S$ will be denoted by $\Gra(\S)$. This is
our set of labels.

As we saw in section 2, the set of labels must be a partially ordered,
directed set. On our set of graphs, the partial order, $\ge$, is
defined just by the inclusion relation:
\be
\g'\ \ge\ \g
\ee
whenever each edge of $\g$ can be expressed as a composition of edges
of $\g'$ {\it and} each vertex in $\g$ is a vertex of $\g'$.

To see that the set is directed, we use the analyticity of edges: it
is easy to check that, given any two graphs $\g_1, \g_2\in \Gra(\S)$,
there exists $\g\in \Gra(\S)$ such that
\be
\g\ge\g_1\ \ \ {\rm and}\ \ \ \g\ge\g_2.
\ee
(In fact, given $\g_1$ and $\g_2$, there exists a {\it minimal} upper
bound $\g$.)  This property is no longer satisfied if one weakens the
definition and only requires that the edges be smooth. 

\subsection{The projective family for $\A$.}

We are now ready to introduce our first projective family.

Fix a graph $\g\in\Gra(\S)$.  To construct the corresponding space
$\A_\g$ in the projective family, restrict the bundle $B$ to the
bundle over $\g$, which we will denote by $B_{\g}$. Clearly, $B_{\g}$
is the union of smooth bundles $B_e$ over the edges of $\g$, $B_{\g}\
=\ \bigcup_{e\in\g} B_{e}$. For every edge $e\in \g$, any connection
$A\in\A$ restricts to a smooth connection $A_e$ on $B_e$. The
collection $(A_e)_{e\in\g} =: A_{|\g}$ will be referred to as the
restriction of $A$ to $\g$. Denote by $\widehat{\G}^\g$ the subgroup
of $\G$ which consists of those vertical automorphisms of $B$ which
act as the identity in the fibers of $B$ over the vertices of $\g$.
Now, since the action of $\G$ on $\A$ is equivariant with the
restriction map $\A\ \rightarrow\ \A_{|\g}$, we can define the
required space $\A_{\g}$ as follows:
\be\label{quo}
\A_\g\ :=\ (\A/\widehat{\G}^{\g})_{\g}.
\ee
Note that $\A_\g$ naturally decomposes into the Cartesian product
\be
\A_\g\ =\ \times_{e\in\g} \A_e
\ee
where $\A_e$ is defined by replacing $\g$ in (\ref{quo}) with a single
edge $e$. Next, let us equip $\A_\g$ with the structure of a
differential manifold.  Note first that, given an orientation of $e$,
a component $A_e$ of $(A_e)_{e\in \g}=A_\g\in\A_\g$, may be identified
with the parallel transport map along the edge $e$ which carries the
fibre over its initial vertex into the fiber over its final vertex.
Hence, if we fix over each vertex of $\g$ a point in $B$ and orient
each edge of $\g$, we have natural maps
\be \label{diff}
\Lambda_e\  :\ \A_e\ \rightarrow\ G, \quad{\rm and}\quad 
\Lambda_\g\ :\ \A_\g \rightarrow\ G^E,
\ee
where $E$ is the number of edges in $\g$. The map can easily be shown
to be a bijection. We shall refer to $\Lambda_e$ (or $\Lambda_\g$) as a
group valued chart for $\A_e$ or ($\A_\g$). Now, since $G$ is a
compact, connected Lie group, $G^E$ is a compact, Hausdorff, analytic
manifold.  Hence, the map $\Lambda_\g$ can be used to endow $\A_\g$
with the same structure.

Finally, we introduce the required projection maps. Note first that,
for each $\g\in \Gra(\S)$, there is a natural projection map
$\pi_{\g}$
\be \label{proj2}
\pi_{\g}\:\ \A\ \rightarrow \ \A_{\g }.
\ee
defined by (\ref{quo}), which is surjective. We now use this map to
define projections $p_{\g\g'}$ between the members of our projective
family. Let $\g'\ \ge\ \g$.  Then, we set
\be
p_{\g\g'}\ :\ \A_{\g'}\ \rightarrow \A_\g
\ee  
to be the map defined by
\be \label{proj1}
\pi_\g\ =\ p_{\g\g'} \circ \pi_{\g'}\ .
\ee
We now have:

\begin{proposition}{\rm :}

\noindent
(i) The map $\Lambda_\g$ of (\ref{diff}) is bijective and the
analytic manifold structure defined on $\A_\g$ by $\Lambda_\g$ does
not depend on the initial choice of points in the fibers of $B$ over
the vertices of $\g$ and orientation of the edges made in its definition;

\noindent
(ii) For every  pair of graphs $\g,\g'\in\Gra(\S)$ such that 
$\g'\ge\g$, the map $p_{\g\g'}$ defined by (\ref{proj2}) and 
(\ref{proj1}) is surjective;  

\noindent
(iii) For any three graphs $\g,\g',\g''\in\Gra(\S)$ such that
$\g''\ge\g'\ge\g$,
\be 
p_{\g\g''}\ =\ p_{\g\g'}\circ p_{\g'\g''} \ ; and,
\ee

\noindent
(iv) The maps $p_{\g\g'}$ are analytic.
\end{proposition}

The proofs are straightforward. It worth noting however, that to show
the surjectivity in (i), one needs the assumption that the
structure group $G$ is connected \cite{AL1}. On the other hand,
compactness of $G$ is not used directly in proposition 1. Compactness
is used, of course, in concluding that $\A_{\g}$ is compact.

Thus, we have introduced a projective family $(\A_\g,
p_{\g\g'})_{\g,\g'\in \Gra(\S)}$ of compact, Hausdorff, analytic
manifolds, labelled by graphs in $\S$. We will denote its projective
limit by $\Ab$.

We will conclude this subsection by presenting a characterization of
$\Ab$. Note first that a connection $A\in\A$ naturally defines a point
$A_\g \in \A_\g$ for each $\g\in\Gra(\S)$ and that the resulting
family $(A_\g)_{\g\in\Gra(\S)}$ represents a point in $\Ab$.  Hence,
we have a natural map
\be
\A\ \rightarrow \ \Ab,
\ee
which is obviously an injection. There are however elements of $\Ab$
which are not in the image of this map. In fact, ``most of'' $\Ab$
lies outside $\A$. To represent a general element of $\Ab$, we proceed
as follows. Consider first a map $I$ which assigns to each oriented
edge $e$ in $\S$, an isomorphism
\be
I(e)\ :\ B_{e_-}\ \rightarrow \ B_{e_+},
\ee
between the fibers $B_{e_\pm}$ over $e_\pm$, the final and the initial
end points of $e$.  Suppose, that this map $I$ satisfies the following
two properties:
\be \label{24}
I(e^{-1}) \ =\ [I(e)]^{-1}\quad {\rm and}\quad I(e_2\circ e_1)\ =\ 
I(e_2)\circ I(e_1)\  ,
\ee
whenever the composed path $e_2\circ e_1$ is again analytic. (Here
$e^{-1}$ is the edge obtained from $e$ by inverting its orientation,
and, if $e_{1+}=e_{2-}$, $e_2\circ e_1$ is the edge obtained by gluing
edges $e_2, e_1$.)  Then, we call $I$ a {\it generalized parallel
transport} in $B$. Let us denote the space of all these generalized
parallel transports by $\P(B)$.  Every element of the projective limit
$\Ab$ defines uniquely an element $I_{\bar A}$ of $\P(B)$. Indeed, let
${\bar A} = (A_\g)_{\g\in\Gra(\S)} \in \Ab$. For an oriented edge $e$
in $M$ pick any graph $\g$ which contains $e$ as the product of its
edges (for some orientation) and define
\be\label{I}
I_{{\bar A}}(e)\ :=\ H(A_\g, e)\   ,
\ee
where the right hand side stands for the (ordinary) parallel transport
defined by $A_\g \in \A_{\g}$.  From the definition of the projective
limit $\Ab$ it is easy to see that (\ref{I}) gives rise to a
well-defined map
\be \label{Hol}
\Ab \ni {\bar A}\ \mapsto\ I_{{\bar A}}\in\P(B) \ .
\ee
Furthermore, it is straightforward to show the following properties of
this map:

\begin{proposition}{\rm :}
The map (\ref{Hol}) defines a one-to-one correspondence between the
projective limit $\Ab$ and the space $\P(B)$ of generalized parallel
transports in $B$.
\end{proposition}

This characterization leads us to regard $\Ab$, heuristically, as the
configuration space of all possible ``floating '' lattices in $\S$,
prior to the removal of gauge freedom at the vertices (see
(\ref{quo})).

\subsection{The projective family for $\G$.}

As we just noted, in the projective family constructed in the last
section, there is still a remaining gauge freedom: given a graph $\g$,
the restrictions of the vertical automorphisms of the bundle $B$ to
the vertices of $\g$ still act non-trivially on $\A_{\g}$.  In this
sub-section, we will construct a projective family $(\G_{\g},
p_{\g\g'})$ from these restricted gauge transformations. In the next
section, we will use the two families to construct the physically
relevant quotient projective family.

Given a graph $\g$, the restricted gauge freedom is the image of the 
following projection
\be \label{proj3}
\tilde{\pi}_\g\ :\ \G\ \rightarrow\ \G/\widehat{\G}^{\g}\ =:\ \G_\g\ .
\ee
Clearly, the group $\G_\g$ has a natural action on $\A_\g$. Since
$\G_\g$ consists essentially of the gauge transformations ``acting at
the vertices'' of $\g$, (up to the natural isomorphism) one can write
$\G_\g$ as the cartesian product group
\be
\G_\g\ =\ \times_{v\in\Ver(\g)}\G_v \ ,
\ee
where $\G_v$ is, as before, the group of automorphisms of the fiber
$\pi^{-1}(v)\subset B$ and $\Ver(\g)$ stands for the set of the
vertices of $\g$. Now, each group $\G_v$ is isomorphic with the
structure group $G$. Hence, if we fix a point in the fiber over each
vertex of $\g$, we obtain an isomorphism
\be \label{diff2}
\tilde{\Lambda}_\g\ :\ \G_\g\  \rightarrow\ G^V\ ,
\ee
where $V$ is the number of edges of $\g$. Finally, given any two graphs
$\g'\ge\g$, the map $\tilde{\pi}_{\g}$ of (\ref{proj3}) factors into 
\be \label{proj4}
\tilde{\pi}_{\g} \ =\ p_{\g\g'}\circ \tilde{\pi}_{\g'}, \ \ \  
p_{\g\g'}\ :\ \G_{\g'}\ 
\rightarrow\ \G_\g, 
\ee
and hence defines the maps $p_{\g\g'}$ uniquely. It is easy to verify
that this machinery is sufficient to endow $(\G_\g, p_{\g\g'})_{\g,\g'
\in \Gra(\S)}$ the structure of a compact, connected Lie group projective
family. We have:

\begin{proposition}{\rm :}

\noindent(i) The family $(\G_\g, p_{\g\g'})_{\g,\g'\in\Gra(\S)}$ 
defined by (\ref{proj3}) and (\ref{proj4}) is a smooth projective
family;

\noindent(ii) the maps $p_{\g\g'}$ are Lie group homomorphisms; 

\noindent(iii) The projective limit $\Gb$ of the family is a compact 
topological  group with respect to the pointwise multiplication: let 
$(g_\g)_{\g\in\Gra(\S)}, 
(h_\delta)_{\delta\in\Gra(\S)} \in \Gb$, then
\be
(g_\g)_{\g\in\Gra(\S)} (h_\delta)_{\delta\in\Gra(\S)}\ :=\ 
(g_\g h_\g)_{\g\in\Gra(\S)}.
\ee

\noindent(iv) There is a natural topological group isomorphism  
\be
\Gb \ \rightarrow \ \times_{x\in\S}\G_x
\ee 
where the group on the right hand side is equipped with the product
topology.
\end{proposition}

In view of the item (iv) above, we again have the expected embedding:
\be
\G \ \rightarrow \Gb\ ,
\ee
where the group $\G$ of automorphisms of $B$ is identified with the
subgroup consisting of those families $(g_x)_{x\in\S}\in\Gb$ which are
smooth in $x$.

Let us equip $\G_\g = G^V$ with the measure $\mu_\g= (\mu_o)^V$, where
$\mu_o$ is the Haar measure on $\G$. Then it is straightforward to
verify that $(\mu_\g)_{\g\in\Gra(\S)}$ is a consistent family of
measures in the sense of section 2. Hence, it defines a regular, Borel
Probability measure $\bar{\mu}_o$ on $\Gb$. This is the just Haar
measure on $\Gb$. Thus, by enlarging the group $\G$ to $\Gb$, one can
obtain a compact group of generalized gauge transformations whose
total volume is {\it finite} (actually, unit). This observation was
first made by Baez \cite{B2}.

\subsection{The quotient $\Ab/\Gb$ and the projective family for $\agb$.}

In the last two subsections, we constructed two projective families.
Their projective limits, $\Ab$ and $\Gb$ are the completions of the
spaces $\A$ and $\G$ of smooth connections and gauge transformations.
The action of $\G$ on $\A$ can be naturally extended to an action of
$\Gb$ on $\Ab$. Indeed, Let $(g_\g)_{\g\in\Gra(\S)}
\in \Gb$ and $(A_\delta)_{\delta\in\Gra(\S)}\in \Ab$. Then, we set
\be
(A_\delta)_{\delta\in\Gra(\S)}\ \ (g_\g)_{\g\in\Gra(\S)} \ :=\ 
(A_\delta g_\delta)_{\delta\in\Gra(\S)}\in\Ab\ ,
\ee
where $(A_\delta, g_\delta)\ \mapsto\ A_\delta g_\delta$ denotes the
action of $\G_\delta$ in $\A_\delta$. Now, this action of $\Gb$ on
$\Ab$ is continuous and $\Gb$ is a compact topological group. Hence,
the quotient, $\Ab/\Gb$, is a Hausdorff and compact space. This concludes
the first part of this subsection.

In the second part, we will examine the spaces $\A_{\g}/\G_{\g}$.
Note first that the projections $p_{\g\g'}$ defined in
(\ref{proj1}), (\ref{proj2}) descend to the projections of the
quotients,
\be \label{proj5}
p_{\g\g'}\ :\ \A_{\g'}/\G_{\g'} \ \rightarrow \ \A_{\g}/\G_{\g}\ .
\ee
We thus have a new compact, Hausdorff, projective family $(\A_\g/\G_\g,
p_{\g\g'})_ {\g,\g'\in\Gra(\S)}$. This family can also be obtained
directly from the quotient $\A/\G$ by a procedure which is analogous
to the one used in section 4.1: The space $\A_{\g}/\G_{\g}$ assigned
to a graph $\g$ is just the image of the restriction map
\be
\pi_\g \ : \ \A/\G\ \rightarrow \ (\A/\G)_{|\g}\ =\ \A_{\g}/\G_{\g}.
\ee
Therefore, it is natural to denote the projective limit of
$(\A_\g/\G_\g, p_{\g\g'})_{\g,\g'\in\Gra(\S)}$ is by
$\overline{\A/\G}$. 

The natural question now is: What is the relation between $\agb$ and
$\Ab/\Gb$? Note first that there is a natural map from $\Ab/\Gb$ to 
$\overline{\A/\G}$, namely
\be \label{iso}
\Ab/\Gb\ni[(A_\g)_{\g\in\Gra(\S)}] \mapsto\   
([A_\g])_{\g\in\Gra(\S)}\in \overline {\A/\G},
\ee
where the square bracket denotes the operation of taking the orbit
with respect to the corresponding group. Using the results of
\cite{MM,AL2}, it is straightforward to show that:

\begin{proposition}
The map (\ref{iso}) defines a homeomorphism with respect to the quotient 
geometry on $\Ab/\Gb$ and the projective limit geometry on 
$\overline{\A/\G}$.
\end{proposition}

Finally, by combining the results of \cite{B1} and \cite{AL2}, one
can show that the space $\agb$ is naturally isomorphic to the Gel'fand
spectrum of the holonomy $C^\star$-algebra $\HAbar$ introduced in
section 1. (Thus, there is no ambiguity in notation.) The space $\agb$
thus serves as the quantum configuration space of the continuum gauge
theory. This is the space of direct physical interest.

We conclude with a number of remarks.
\begin{enumerate}
\item Since $\agb$ is the Gel'fand spectrum of $\HAbar$, it follows 
\cite{AL1} that there is a natural embedding:
\be
\bigcup_{B'}\ (\ag)_{B'}\ \rightarrow \agb,
\ee
where $(\ag)_{B'}$ denotes the quotient space of connections on a bundle
$B'$ and $B'$ runs through all the $G$-principal fibre bundles over
$\S$. Thus, although it is not obvious from our construction, $\agb$
is independent of the choice of the bundle $B$ we made in the
beginning; it is tied only to the underlying manifold $\S$. (See
\cite{AI,AL1,B1,MM} for the bundle independent definitions.)
\item Each member $\A_\g$ and $\G_\g$ of the first two projective 
families, we considered is a compact, analytic manifold.
Unfortunately, the same is not true of the quotients $\A_\g/\G_\g$
which constitute the third family since the quotient construction
introduces kinks and boundaries.  Because of this, while discussing
differential geometry, we will regard $\agb$ as $\Ab/\Gb$ and deal
with $\Gb$ invariant structures on $\Ab$. That is, it would be more
convenient to work ``upstairs'' on $\Ab$ even though $\agb$ is the
space of direct physical interest. This point was first emphasized by
Baez \cite{B1, B2}.

\item In the literature, one often fixes a base point $x_0$ in $\S$ 
and uses the subgroup $\G_{x_0}$ of $\G$ consisting of vertical
automorphisms which act as the identity on the fiber over $x_0\in\S$
as the gauge group. In the present framework, this corresponds to
considering the subgroups $\G_{\g,x_0}\subset\G_{\g}$ where $\g$ run
through $\Gra(\S)_{x_0}$, the space of graphs which have $x_0$ as a
vertex. $\agb$ can be recovered by taking the quotient of the
projective limit of this family by the natural action of the gauge
group at the base point.
\end{enumerate}

\section{Elements of the differential geometry on $\Ab$}

We are now ready to discuss differential geometry. We saw in section 2
that one can introduce a measure on the projective limit by specifying
a consistent family of measures on the members of the projective
family. The idea now is to use this strategy to introduce on the
projective limits various structures from differential geometry.  The
object of our primary interest is $\agb$. However, as indicated above,
we will first introduce geometric structures of $\Ab$.  Those
structures which are invariant under the action of $\Gb$ on $\Ab$ will
descend to $\agb = \Ab/\Gb$ and provide us, in section 5, with
differential geometry on the quotient.

In section 4.1, we introduce $C^n$ differential forms on $\Ab$ and, in
section 4.2, the $C^n$ volume forms. Section 4.3 is devoted to $C^n$
vector fields and their properties. Finally, in section 4.4, we
combine these results to show how vector fields can be used to define
``momentum operators'' in the quantum theory. While we will focus on
the projective family introduced in section 3.2, our analysis will go
through for {\it any} projective family, the members of which are
smooth compact manifolds.

Throughout this section $C^n$ could in particular stand for $C^\infty$
or $C^\omega$. 

\subsection{Differential forms}

Let us begin with functions. 

Results of section 2 imply that the projective limit $\Ab$ of the
family $(\A_\g, p_{\g \g'})_{\g,\g'\in\Gra(\S)}$ is a compact
Hausdorff space. Hence, we have a well-defined algebra $C^0(\Ab)$ of
continuous functions on $\Ab$. We now want to introduce the notion of
$C^n$ functions on $\Ab$. The problem is that $\Ab$ does not have a
natural manifold structure. Recall however that the algebra $C^0(\Ab)$
could also be constructed directly from the projective family, without
passing to the limit: We saw in section 2 that $C^0(\Ab)$ is naturally
isomorphic with the algebra $\Cyl^0 (\Ab)$ of cylindrical continuous
functions. The idea now is to simply define differentiable functions
on $\Ab$ as cylindrical, differential functions on the projective
family.

This is possible because each member $\A_\g$ of the family has the
structure of an analytic manifold, and the projections $p_{\g\g'}$ are
all analytic. Thus, we can define $C^n$ cylindrical functions
$\Cyl^n(\Ab)$ to be
\be
\Cyl^n(\Ab)\ :=\ \bigcup_{\g\in\Gamma}C^n(\A_\g)/\sim,
\ee
where the equivalence relation is the same as in (\ref{2.6}) of
section 2; as before, it removes the redundancy by identifying, if
$\g'\ge \g$, the function $f$ on $\A_\g$ with its pull-back $f'$ on
$\A_{\g'}$. Elements of $\Cyl^n(\Ab)$ will serve as the $C^n$
functions on $\Ab$.  Note that if a cylindrical function $f\in
\Cyl(\Ab)$ can be represented by a function $f_\g\in C^n(\A_\g)$, then
all the representatives of $f$ are of the $C^n$ differentiability
class.


Next we consider higher order forms. The idea is again to use an
equivalence relation $\sim$ to ``glue'' differential forms on
$(\A_\g)_{\g\in\Gra(\S)}$ and obtain strings that can serve as
differential forms on $\Ab$.  Consider $\bigcup_{\g\in\Gamma}\Omega
(\A_\g)$, where $\Omega (\A_\g)$ denotes the Grassman algebra of all
$C^n$ sections of the bundle of differential forms on $\A_\g$.  Let us
introduce the equivalence relation $\sim$ by extending (\ref{2.6}) in
an obvious way:
\be \label{62}
\Omega(\A_{\g_1}) \ni \omega_{\g_1}\ \sim\ \omega_{\g_2}\in \Omega
(\A_{\g_2})\ \ \  {\rm iff} \ \ \ p_{\g_1\g'}^\star\omega_{\g_1}\  
= \ p_{\g_2\g'}^\star\omega_{\g_2},
\ee
for any $\g'\ge\g_1,\g_2$. (Again, if the equality above is true for a
particular $\g'$ then it is true for every $\g'\ge \g_1, \g_2$.)  The
set of differential forms on $\Ab$ we are seeking is now given by:
\be
\Omega(\Ab)\ :=\  (\bigcup_{\g\in\Gamma}\Omega(\A_\g))\ / \sim.
\ee 
Clearly, $\Omega(\Ab)$ contains well-defined subspaces $\Omega^m(\Ab)$
of $m$--forms. Since the pullbacks $p_{\g\g'}^\star$ commute with the
exterior derivatives, there is a natural, well-defined exterior
derivative operation $d$ on $\Ab$:
\be
d\ : \Omega^n(\Ab)\ \rightarrow\ \Omega^{n+1}(\Ab).
\ee
One can use it to define and study the corresponding cohomology groups
$H^n(\Ab)$. 

Thus, although $\Ab$ does not have a natural manifold structure, using
the projective family and an algebraic approach to geometry, we can
introduce on it $C^n$ differential forms and exterior calculus.

\subsection{Volume forms.}

Volume forms require a special treatment because they are not
encompassed in the discussion of the previous section. To see this,
recall that an element of $\Omega(\Ab)$ is an assignment of a
consistent family of $m$-forms, for some {\it fixed} $m$, to each
$\A_\g$, with $\g\ge\g_o$ for some $\g_o$.  On the other hand, since a
volume form on $\Ab$ is to enable us to integrate elements of
$\Cyl^0(\Ab)$, it should correspond to a consistent family of
$d_\g$-forms on $(\A_\g)_{\g\in \Gra(\S)}$, where $d_\g$ is the
dimension of the manifold $\A_\g$. That is, the rank of the form is no
longer fixed but changes with the dimension of $\A_{\g}$. Thus,
volume forms are analogous to the measures discussed in section 2
rather than to the n-forms discussed above.

The procedure to introduce them is pretty obvious from our discussion
of measures. A {\it $C^n$ volume form} on $\Ab$ will be a family
$(\nu_\g)_{\g\in\Gra(\S)}$, where each $\nu_\g$ is a $C^n$ volume form
with strictly positive volume on $\A_\g$, such that
\be \label{volcons}
C^0(\A_{\g'})\ni f_{\g'}\ \sim\ f_\g\in 
C^0(\A_{\g})\ \ \Rightarrow \ \  \int_{\A_\g}\ f_\g\nu_\g =\  
\int_{\A_{\g'}}\ f_{\g'}\nu_{\g'}\ ,
\ee
for all $\g'\ge \g$ and all functions $f_\g$ on $\A_\g$, where
$f_{\g'} = p_{\g \g'}^\star f_\g$.  Now, since $\A_\g$ are all compact,
it follows from the discussion of measures in section 2 that this
volume form automatically defines a regular Borel measure, say $\nu$,
on $\Ab$ and that this measure satisfies:
\be
\int_\Ab\  [f_\g]_\sim\ d\nu\ :=\ \int_{\A_\g}f_\g\nu_\g.
\ee

The most natural volume form $\mu_o$ on $\Ab$ is provided by the
normalized, left and right invariant (i.e., Haar) volume form $\mu_H$
on the structure group $G$.  Use the map $\Lambda_\g:\A_\g\
\rightarrow G^E$ defined in (\ref{diff}) to pull back to $\A_\g$ the
product volume form $(\mu_H)^E$, induced on $G^E$ by $\mu_H$, to
obtain
\be \label{vol}
\mu^H_\g\ :=\ \Lambda_{\g}^{-1}{}_\star\ (\mu_H)^E.
\ee
We then have \cite{AL1,B2,AL2}:

\begin{proposition}{\rm :}

\noindent(i) The form $\mu^H_\g$ of (\ref{vol}) is insensitive
to the choice of the gauge over the vertices of $\g$,
used in the definition (\ref{diff}) of the map $\Lambda_\g$; 

\noindent(ii) The family of volume forms $(\mu^H_\g)_{\g\in\Gra(\S)}$ 
satisfies the consistency conditions (\ref{volcons});

\noindent(iii) The volume form $\mu_o$ defined on $\Ab$ by 
$(\mu^H_\g)_{\g\in \Gra(\S)}$ is invariant with respect to the action
of (all) the automorphisms of the underlying bundle $B(M,G)$.
\end{proposition}

The push forward $\mu_o'$ of $\mu_o$, under the natural projection
from $\Ab$ to $\agb$ is the induced Haar measure on $\agb$ of
\cite{AL1}, chronologically, the first measure introduced in this 
subject. It is invariant with respect to all the diffeomorphisms of 
$M$.

The measure $\mu_o$ itself was first introduced in \cite{B1}.  By now,
several infinite families of measures have been introduced on $\Ab$
(which can be pushed forward to $\agb$) \cite{B1,B2,ALMMT2,ALMMT1}.
These are reviewed in \cite{AL2}. In section 6, using heat kernel
methods, we will introduce another infinite family of measures.
These, as well as the measures introduced in \cite{AL2} arise
from $C^{n}$ volume forms on $\Ab$.

\subsection{Vector fields.}

Introduction of the notion of vector fields on $\Ab$ is somewhat more
subtle than that of $m$-forms because while one can pull-back forms,
in general one can push forward only vectors (rather than vector
fields). Hence, given $\g'\ge\g$, only certain vector fields on
$\A_{\g'}$ can be pushed-forward through $(p_{\g\g'})^\star$. To
obtain interesting examples, therefore, we now have to introduce an
additional structure: vector fields on $\Ab$ will be associated with a
graph.

A {\it smooth vector field $X^{(\g_o)}$ on $\Ab$} is a family
$(X_\g)_{\g\ge\g_o}$ where $X_\g$ is a smooth vector field on $\A_\g$
for all $\g\ge \g_o$, which satisfies the following consistency
condition:
\be \label{vecons}
(p_{\g\g'})_{\star}\ X_{\g'}\ =\ X_\g,\ \ \ {\rm whenever}\ \ \
\g'\ge\g\ge \g_0.
\ee
It is natural to define a derivation $D$ on $\Cyl^n(\Ab)$, as a linear
and star preserving map, $D\:\ \Cyl^n(\Ab)\ \rightarrow\ \Cyl^{n-1}
(\Ab)$, such that for every $f,g\in \Cyl^n(\Ab)$, the Leibniz rule
holds, i.e., $D(fg)\ =\ D(f)g + D(g)f$.  As one might expect, a vector
field $X^{(\g_o)}$ defines a derivation, which we will denote also by
$X^{(\g_o)}$.  Indeed, given $f\in\Cyl^n(\Ab)$ there exists
$\g\ge\g_0$ such that $f\ =\ [f_\g]_\sim$. We simply set
\be
 X^{(\g_o)}(f)\:=\ [X_\g(f_\g)]_\sim\in\Cyl^{n-1}(\Ab),
\ee  
where $X_\g(f_\g)$ is the action of the vector field $X_\g$ on
$\A_\g$ on the function $f_\g$, and note that the right hand side is
independent on the choice of the representative.

Finally, given any two vector fields, we can take their commutator.
We have:
\begin{proposition}{\rm :} Let $X^{(\g_1)} = (X_\g)_{\g\ge\g_1}$, 
$Y^{(\g_2)} = (Y_\g)_{\g\ge\g_2}$ be two vector fields on $\Ab$. Then,
the commutator $[X^{(\g_1)}, Y^{(\g_2)}]$ of the corresponding
derivations is the derivation defined by a vector field $Z^{(\g_3)}$
on $\Ab$ (where $\g_3\in \Gra(\S)$ is any label satisfying $\g_3 \ge
\g_1,\g_2$) given by
\be
 Z_\g\ =\ [X_\g,Z_\g],
\ee 
for any $\g\ge \g_3$.
\end{proposition} 

For notational simplicity, from now on, we will drop the superscripts
on the vector fields.

\bigskip\noindent

\subsection{Vector fields as momentum operators}

We will first introduce the notion of compatibility between vector
fields $X$ and volume forms $\mu$ on $\Ab$ and then use it to define 
certain essentially self-adjoint operators $P(X)$ on $L^2(\Ab,\mu)$.

Let us begin by recalling that, given a manifold $\Sigma$ a vector
field $V$ and a volume form $\nu$ thereon, the divergence $\div_\nu V$
is a function defined on $M$ by
\be
L_V\ \nu\ =:\ (\div_\nu V) \nu,
\ee
where $L_V$ denotes the standard Lie derivative. 

We will say that a vector field $X=(X_\g)_{\g\ge\g_0}$ on $\Ab$ is
{\it compatible with a volume form} $\mu=(\mu_\g)_{\g\in
\Gra(\S)}$ on $\Ab$ if 
\be \label{div}
p_{\g\g'}^\star\ \div_{\mu_\g} X_{\g}\ =\ \div_{\mu_{\g'}}\ 
X_{\g'},
\ee    
whenever $\g'\ge\g\ge\g_0$.  Note, that if (\ref{div}) holds, the
divergence $\div_{\mu_\g}\ X_\g$ is a cylindrical function,
\be
\div_\mu X\:=\ [\div_{\mu_\g}\ X_\g]_\sim\in\Cyl^\infty(\Ab).
\ee
We shall call it {\it the divergence of} $X$ with respect to a volume
form $\mu$. The next proposition shows that the divergence of vector
fields on $\Ab$ has several of the properties of the usual, finite
dimensional divergence.

\begin{proposition}{\rm :}

\noindent i) Let $X$ be a vector field and $\mu$, a smooth volume
form on $\Ab$ such that $X$ is compatible with $\mu$. Then,
for every $f,g\in\Cyl^1(\Ab)$,
\be
\int_\Ab f X(g)\mu \ =\ -\int_\Ab (X(f)+ (\div_\mu\ X) f)g\ \mu.
\ee

\noindent ii) Suppose that $Y$ is another vector field on $\Ab$ which
is compatible with $\mu$. Then, the commutator $[X,Y]$ also is
compatible with $\mu$, and
\be
\div_\mu [X, Y]\ =\ X(\div_\mu\ Y)\ -\ Y(\div_\mu\  X). 
\ee
\end{proposition}

{\it Proof}{\rm :} The result follows immediately by using the
properties of the usual divergence of vector fields $X_\g$ and $Y_\g$
on $(\A_\g,\mu_\g)$ and the consistency conditions satisfied by $X_\g,
Y_\g$ and $\mu_\g$.

$\Box$

\bigskip 

We are now ready to introduce the momentum operators.  Fix a smooth
volume form $\mu \ =\ (\mu_\g)_{\g\in \Gra(\S)}$ on $\Ab$.  In the
Hilbert space $L^2(\Ab,\mu)$, we define below a quantum representation
of the Lie algebra of vector fields compatible with $\mu$. Let $X$ be
such a vector field on $\Ab$. We assign to $X$ the operator $(P(X),
\Cyl^1(\Ab))$ as follows:
\be \label{P(X)}
P(X)\ := \ iX\ +\ {i\over 2} (\div_\mu\ X)\ .
\ee
(Here, $\Cyl^1(\Ab)$ is the domain of the operator.) Clearly, $(P(X), 
\Cyl^1(\Ab))$ is a densely-defined operator on the Hilbert space
$L^2(\Ab, \mu)$.  Following the terminology used in quantum mechanics
on manifolds, we will refer to $P(X)$ as the momentum operator
associated with the vector field $X$. As one might expect from this
analogy, the second term in the definition (\ref{P(X)}) of the
momentum operators is necessary to ensure that it is a symmetric
operator.

To examine properties of this operator, we first need some general
results. Let us therefore make a brief detour and work in a more
general setting. Consider a family of Hilbert spaces $(\H_\g, p_{\g
\g'}^\star)_{\g, \g' \in \Gamma}$ where $\Gamma$ is any partially ordered
and directed set of labels and
\be \label{projh}
p_{\g \g'}^\star:\ \H_\g\ \rightarrow\ \H_{\g'}.  
\ee 
is an inner-product preserving embedding defined for each ordered pair
$\g'\ge\g\in\Gamma$. The maps (\ref{projh}) provide the union
$\bigcup_{\g\in\Gamma}H_\g$ with an equivalence relation defined as in
(\ref{2.6}). The Hermitian inner products $(.,.)_\g$ give rise to a
unique Hermitian inner product on the vector space $(\bigcup_{\g\in
\Gamma} H_\g) /\sim$ . For, if $\psi,\phi\in (\bigcup_{\g\in\Gamma}
H_\g)/\sim$ , there exists a common label $\g\in\Gamma$ such that
$\psi\ =\ [\psi_\g]_\sim$ and $\phi\ =\ [\phi_\g]_\sim$, with $\psi_\g,
\phi_\g\in H_\g$, and we can set
\be
(\psi, \phi)\ :=\ (\psi_\g,\phi_\g)_\g.
\ee
It is easy to check that this inner product is Hermitian.  Thus, we
have a pre-Hilbert space. Let $\H$ denote its Cauchy completion:
\be
\H\ =\  \overline{\bigcup_{\g\in\Ga}\H_\g/\sim}.
\ee

On this Hilbert space $\H$, consider an operator given by a family of
operators $(O_\g, \D_{\g}(O_\g))_{\g\in\Ga(O)}$, where
$\Ga(O)\subset\Gamma$ is a cofinal subset of labels (i.e., for every
$\g\in\Gamma$ there is $\g'\in\Gamma(O)$ such that $\g'\ge\g$.) We
will say that $(O_\g, \D_{\g}(O_\g))_{\g\in\Ga(O)}$ is self consistent
if the following two conditions are satisfied:
\be \label{01}
p_{\g\g'}^\star\D_\g(O_\g)\ \subseteq\ \D_{\g'}(O_{\g'})
\ee

\be \label{02}
O_{\g'}\circ\ p_{\g\g'}^\star \  = \ p_{\g\g'}^\star\circ O_\g
\ee
for every $\g'\ge\g$ such that $\g',\g\in\Ga(O).$ Since the label set
$\Ga(O)\subset \Ga$ is cofinal, a self consistent family
of operators $(O_\g, \D_{\g}(O_\g))_{\g \in\Ga(\P)}$ defines an
operator $O$ in $\H$ via $O(\psi)\ :=\ [O_\g \psi_\g]_\sim$.

A general result which we will apply to the momentum operators 
is the following.

\begin{lemma}{\rm :} \label{selfadjointness}
Let $(O_\g, \D_{\g}(O_\g))_{\g\in\Ga(O)}$ be a self consistent family
of operators and $\Ga(O)$ be cofinal in $\Ga$. Then,

\noindent(i) $(O_\g, \D_{\g}(O_\g))_{\g\in\Ga(\P)}$ defines
uniquely an operator $O$ in $\H$ acting on a domain $\D(O) \:=\
\bigcup_{\g\in\Ga(O)}{\D_\g(O_\g)}/\sim$ and such that for every
$f_\g\in\D_\g(O_\g)$
\be
O([f_\g]_\sim)\ =\ [O_\g(f_\g)]_\sim;
\ee

\noindent(ii) If $(O_\g, \D_\g(O_\g))$ is essentially self-adjoint 
in $H_\g$ for every $\g\in \Ga(O)$, then the resulting operator $(O,
\D(O))$ defined in (i) is also essentially self adjoint;

\noindent(iii) If $(O_\g, \D_\g(O_\g))$ is essentially 
self-adjoint in $H_\g$ for every $\g\in \Ga(O)$, then the family of
the self dual extensions $({\tilde O}_\g, \D_\g({\tilde 
O}_\g))_{\g\in\Ga(O)}$ is self consistent.
\end{lemma}
\bigskip
{\it Proof}{\rm :} Part (i) is obvious from the above discussion. 

We will prove (ii) by showing that the ranges of the operators $O+iI$
and $O-iI$, where $I$ is the identity operator, are dense in $\H$.
They are given by
\be \label{range}
(O\pm iI) ( \D(O))\ =\  \bigcup_{\g\in
\Ga(O)}(O_\g\pm i I)(\D_\g(O_\g))/\sim.
\ee
But, as follows from the hypothesis, the range of each of the
operators $O_\g\pm i I$ is dense in the corresponding $H_\g$. Hence
indeed, the right hand side of (\ref{range}) is dense in $\H$.

To show (iii), recall that the self-adjoint extension of an
essentially self-adjoint operator is just its closure.  Let
$\g',\g\in\Ga(O)$ and $\g'\ge\g$. Via the pullback $p_{\g\g'}^\star$,
we may consider $H_{\g}$ as a subspace of $\H_{\g'}$. Since
$(O_{\g'},\D_{\g'}(O_{\g'})$ is an extension of $(O_\g,
\D_{\g}(O_{\g}))$, the closure $({\tilde O}_{\g'}, \D_{\g'}
({\tilde O}_{\g'}))$ is still an extension for $({\tilde O}_{\g},
\D_{\g}({\tilde O}_{\g}))$. This concludes the proof of the Lemma.

$\Box$

\bigskip

We can now return to the momentum operators $(P(X), \Cyl^1(\Ab))$
on $L^2(\Ab, \mu)$. 

\begin{theorem}{\rm :} 

\noindent
(i)Let $X =(X_\g)_{\g\ge\g_0}$ and $\mu=(\mu_\g)_{\g\in\Gra(\S)}$ be a
smooth vector field and volume form on the projective limit $\Ab$.
Suppose $X$ is compatible with $\mu$; then,  the operator
$(P(Y), \Cyl^1(\Ab))$ of (\ref{P(X)}) is essentially self-adjoint on
$L^2(\Ab,\mu)$;

\noindent 
(ii) Let $Y$ be another smooth vector field on the projective 
limit, also compatible with the measure $\mu$. Then, the vector field 
$[X,Y]$ also is compatible with $\mu$ and 
\be
P([X,Y])\ = \ i [P(X),P(Y)].
\ee  
\end{theorem}

{\it Proof}{\rm:}
Part $(i)$ of the Theorem follows trivially from Lemma 1; we only have
to substitute $L^2(\Ab,\mu)$ for $\H$, $L^2(\A_\g, \mu_\g)$ for
$H_\g$, $\Gra(\S)$ for $\Gamma$ and $((i(Y_\g+{1\over
2}\div_{\mu_\g}Y_\g))_{\g \ge\g_0}, \\
\Cyl^1(\Ab))$ for $(O_\g, \D_{\g}(O_\g) )_{\g\in\Ga(O)}$. 

Finally, part $(ii)$ can be shown by a simple calculation using
Proposition 7.  

$\Box$

\bigskip

This concludes our discussion of the momentum operators.  Most of the
results of this section concern the case when a vector field is
compatible with a volume form. In the next section we shall see that a
natural symmetry condition implies that a vector field on $\Ab$ is
necessarily compatible with the Haar volume form.

\section{Elements of differential geometry on $\agb$}

We now turn to $\agb$, the space that we are directly interested in.

\subsection{Forms and volume forms}

Let us begin with functions. We know from section 3.4 that 
\be
\overline{\A/\G}\ =\ \Ab/\Gb\ .
\ee
Therefore, we can drop the distinction between functions on
$\overline{\A/\G}$ and $\Gb$-invariant functions defined on $\Ab$. In
particular, we can identify the $C^\star$-algebra $C^0(\agb)$ of
continuous functions on $\agb$ with the $C^\star$-subalgebra of $\Gb$
invariant elements of $C^0(\Ab)$. This suggests that we adopt the same
strategy towards differentiable functions and forms.  Therefore, we
will let $\Cyl^n(\agb)$ be the $\star$-subalgebra of $\Gb$-invariant
elements of $\Cyl^n(\A)$, and $\Omega(\agb)$ be the sub-algebra
consisting of $\G$-invariant elements of Grassmann algebra
$\Omega(\Ab)$. The operations of taking exterior product and exterior
derivative is well-defined on $\Omega(\agb)$. 

Similarly, by a volume form on $\agb$, we shall mean a $\G$-invariant
volume form on $\Ab$. As noted in section 4.2, the induced Haar
form on $\Ab$ is $\Gb$ invariant and provides us with a natural
measure on $\agb$. Furthermore, since $\Gb$ is compact, we can 
extract the $\Gb$-invariant part of {\it any} volume form $\nu$ on
$\Ab$ by an averaging procedure (see also \cite{B1}).

\begin{proposition}{\rm :}
Suppose $\nu\ =\ (\nu_\g)_{\g\in \Gra(\S)}$ is a volume form on $\Ab$.
Then, if $R_{\bar g}$ denotes the action of $\bar{g}\in \Gb$ on $\Ab$,
and $d\bar g$ denotes the Haar measure on $\Gb$,
\be
\overline {\nu}\ :=\ \int_\Gb\ (\R_{{\bar g})_\star}\nu\  d{\bar g},
\ee
is a $\Gb$ invariant volume form on $\Ab$ such that for every
$\Gb$ invariant function  $f\in C^0(\Ab)$
\be
\int_\Ab\ f\ \nu\ =\ \int_\Ab\ f\ \overline{\nu}\  .
\ee  
\end{proposition}

In terms of the projective family, we can write out the averaged
volume form more explicitly. Let $\nu\ =\ (\nu_{\g})_{\g\in
\Gra(\S)}$.  Then, the averaged volume-form $\overline {\nu}\ =\
(\overline{\nu_\g})_{\g\in\Gra(\S)}$ where
\be
\overline{\nu_\g}\ =\ \int_{G^V}\ (\R_{(g_1,...,g_V)})_\star\nu_\g
\ \ dg_1\wedge...\wedge dg_n\ ,
\ee
where $\R_{(g_1,...,g_V)}$ denotes the action of $(g_1,...,g_V)\in
\G_\g $ on $\A_\g$ ($\G_\g$ being identified with
$G^V$).

\subsection{Vector fields on $\agb$}

The procedure that led us to forms on $\agb$ can also be used to
define vector fields on $\agb$.  Furthermore, for vector fields, one
can obtain some general results which are directly useful in
defining quantum operators. We will establish these results in this
subsection and use them to obtain a complete characterization of
vector fields on $\agb$ in the next subsection.

The group $\Gb$ acts on vector fields on $\Ab$ as follows. Let
$X=(X_\g)_{\g\ge\g_0}$, and ${\bar g}=(g_\g)_{\g\in\Gra(\S)}\in\Gb$.
Then
\be
R_{\bar g}\star\ X\ :=\ (R_{g_\g\star}\ X_\g)_{\g\ge\g_0}\ ,
\ee
where, as before, $R_{g_\g}:\A_\g\rightarrow\A_\g$ is the action of
$\G_\g$ on $\A_\g$. In the first part of this subsection, we will
explore the relation between $\Gb$-invariant vector fields and the
induced Haar form $\mu_o$ on $\Ab$. Since we are dealing here only
with the Haar measure, for simplicity of notation, in this
subsection, we will drop the (measure-)suffix on divergence.

\begin{theorem}\label{div2}{\rm :}  Let $X=(X_\g)_{\g\ge\g_0}$ be a $C^{n+1}$ 
vector field on $\Ab$. If $X$ is $\Gb$ invariant then it is compatible
with the Haar measure $\mu_o$ on $\Ab$ and $\div\ X\in\Cyl^{n}(\Ab)$
is $\Gb$ invariant.
\end{theorem} 
 
{\it Proof}{\rm :} We need to show that, if $\g, \g' \ge\g_0$, then
\be
\div\ X_{\g} \sim  \div\ X_{\g'}\ .
\ee
Since the family $(X_\g)_{\g\ge\g_0}$ is consistent, it is sufficient
to show that if $\g_2\ge \g_1$, then
\be \label{diveq}
p_{\g_1\g_2}^\star\ (\div\ (p_{\g_1\g_2})_\star\ X_{\g_2}) = \div\ X_{\g_2}\ .
\ee
Now, the graph $\g_2$ consists of two types of edges: i) edges
${e}_1,...,{e_{E_3}}$, say, which are contained in $\g_1$, and ii) the
remaining edges ${e'}_{E_3+1},...,{e'_{E_2}}$, say. 
The first set of edges
forms a graph ${\g_3}\ge \g_1$ whose image coincides with that of
$\g_1$.  Therefore, in particular, we have $\g_3\ge {\g_0}$ and
$X_{\g_3}$ is well-defined. Our strategy is to decompose the
projection $p_{\g_1\g_2}$ as
\be
p_{\g_1\g_2}\ =\  p_{\g_1\g_3}\circ p_{\g_3\g_2}\ .
\ee
and prove that each of the two projections on the right hand side
satisfies (\ref{diveq}), i.e., that we have
\be \label{diveq1}
p_{\g_1\g_3}^\star\ (\div\ (p_{\g_1\g_3})_\star\ X_{\g_3}) = \div\ X_{\g_3}\ ,
\ee
and 
\be \label{diveq2}
p_{\g_3\g_2}^\star\ (\div\ (p_{\g_3\g_2})_\star\ X_{\g_2}) = \div\ X_{\g_2}\ .
\ee
These two results will be established in two lemmas which will
conclude the proof of the main part of the theorem.

Once this part is established, the $\Gb$-invariance of
$\div(X)\in\Cyl^\infty(\Ab)$ is obvious from the $\Gb$-invariance of
the vector field $X$ and of the measure $\mu_o$.

\bigskip

\begin{lemma}{\rm :}
Let $\g_3\ge \g_1$ be such that the images of $\g_3$ and $\g_1$ in
$\S$ coincide. Let $X_{\g_3}$ be a $\G_{\g_3}$-invariant vector field
on $\A_{\g_3}$ and $X_{\g_1}$, a $\G_{\g_1}$-invariant vector field on
$A_{\g_1}$ such that $(p_{\g_1\g_3})_\star X_{\g_3}\ =\ X_{\g_1}$. Then,
\be
p_{\g_1\g_3}^\star (\div X_{\g_1})\ =\ \div X_{\g_3}.
\ee
\end{lemma}
\medskip
{\it Proof}{\rm :} Since $\g_3$ is obtained just by subdividing some
of the edges of $\g_1$, it follows that the pull-back $p_{\g_1\g_3}^\star$
is an isomorphism of the $C^\star$-algebra of continuous and $\G_{\g_1}$
invariant functions on $\A_{\g_1}$ into the $C^\star$-algebra of
continuous and $\G_{\g_3}$ invariant functions on $\A_{\g_3}$. Hence
it defines an isomorphism between the corresponding Hilbert
spaces. The vector fields $X_{\g_1}$ and $X_{\g_3}$ define the
operators which are equal to each other via the isomorphism. Hence,
their divergences are also equivalent as operators, and being smooth
functions, are just equal to each other, modulo the pullback. 
\medskip

Now, we  turn to the second part, (\ref{diveq2}), of the proof.
\begin{lemma}\label{3}{\rm :}
Let $X= (X_\g)_{\g\ge \g_0}$ be a $\G$-invariant 
vector field on $\Ab$. Let $\g_2
\ge\g_3 \ge \g_0$ be such that $\g_2$ is obtained by adding edges 
to $\g_3$ the images of all of which, except possibly the end points, 
lie outside the image of $\g_3$. Then, 
\be
p_{\g_3\g_2}^\star (\div X_{\g_3})\ =\ \div X_{\g_2}\ . 
\ee
\end{lemma}
\medskip
{\it Proof}{\rm :} Via appropriate parametrization we can set 
\be
\A_{\g_2}\ =\ \A_{\g_3}\times G^{E_2-E_3},
\ee
so that the map $p_{\g_3\g_2}$, becomes the obvious projection
\be
p_{\g_3 \g_2}\ :\ \ \A_{\g_3}\times G^{E_2-E_3}\ \rightarrow\ \A_{\g_3}\ .
\ee
Since $X_{\g_2}$ projects unambiguously to $X_{\g_3}$, it follows that
we decompose $X_{\g_2}$ as:
\be
X_{\g_2} = (X_{\g_3}, X_{E_3+1},...,X_{E_2}),
\ee
where, for each choice of a point on $\A_{\g_3}$, and of variables
$g_{E_3+j}$, $j\not= i$, we can regard $X_{E_3+i}$, as a vector field
on $G$. (Here, $i= 1,...,E_2-E_3$, and $\A_{E_3 +k}$ is identified
with $G$).

We will now analyze the properties of these vector fields. Let us fix
an edge $e_{E_3+i}$. We will now show that $X_{E_3+i}$ does not change
as we vary $g_{E_3+1},..., g_{E_3+i-1}, g_{E_3+i+1},..., g_{E_2}$. Let
us suppose that there exist some edges $e_{E_3+j}$ which are {\it
removable} in the sense that one can obtain a closed graph $\g_4\ge
\g_3$ after removing them. Clearly, $\g_2\ge\g_4 \ge \g_0$. Hence, there
is a vector field $X_{\g_4}$ on $\A_{\g_4}$ such that $X_{\g_3},
X_{\g_4}, X_{\g_2}$ are all consistent. This implies that $X_{E_3 +i}$
does not change if we vary $g_{E_3+j}$. Now let us consider the case
when edges of $\g_{3}$ are not removable. Then, we can construct a
closed graph $\g_5$
\be
{\g_5} := {\g_2}\cup\{e_+,e_-\}\ ,
\ee
by adding two new edges $e_{\pm}$ to join the vertices of $e_{E_3 +i}$
to any two vertices of $\g_3$. Then, $\g_5\ge \g_2\ge \g_3$ and we
have consistency of the vector fields $X_{\g_5}, X_{\g_2}, X_{\g_3}$.
Clearly, the $X_{E_3+i}$ component of $X_{\g_5}$ coincides with the
$X_{E_3 +i}$ component of $X_{\g_2}$. But in $X_{\g_5}$, all the
edges $e_{E_3+ j}$ with $j \not= i$ are removable. Hence,
$X_{E_3+i}$ does not depend on $g_{E_3+j}$ if $i\not=j$. Thus, we have
shown that $X_{E_3 +i}$ is independent of $g_j$ if $i\not=j$.

So far, in this lemma, we have only used the consistency of
$(X_\g)_{\g\ge \g_0}$. We now use the $\G$-invariance of $X$ to show
that (for each $g_{\g_3} \in \A_{\g_3}$) $X_{E_3+i}$ is a left
invariant vector field on $G$. Let $v$ be a vertex of $e_{E_3+i}$
which is not contained in $\g_3$. For definiteness, let us suppose that
it is the final vertex. Then, under the gauge transformation $a$
in the fiber over this vertex, we have:
\be 
(a_\star\ X_{\g_2})_{E+i}\ =\ (L_{a^{-1}})_\star\ X_{E+i}\ ,
\ee
where the left side is the $E_{3+i}$-th component of the vector field
in the parenthesis and $L_a$ is the left action of $a$ on $G$.  (Here,
we have used our earlier result that $X_{E+i}$ does not depend on
$g_{E+j}$ when $i\not= j$.) Now, from $\G_{\g_2}$ invariance of
$X_{\g_2}$, it follows that
\be 
X_{E+i}\ =\  {{\rm L}_a}_\star X_{E+i}.
\ee
This conclusion applies to any value of $i=1,...,E_2-E_3$. Thus, for 
each choice of $g_{\g_3}\in \A_{\g_3}$,  $X_{E_3+i}$ are, in particular,
divergence-free vector fields on $G$.

We now collect these results to compute the divergence of $\X_{\g_2}$:
\be
\div X_{\g_2}\ =\ \div X_{\g_3} + \div X_{E_3+1}\ + ...+ \div X_{E_2}\ 
= \ \div X_{\g_3}\ ,
\ee
where, for simplicity of notation, we have dropped the pull-back symbols.

$\Box$

\bigskip

Using this result and those of section 4.4, we have the following
theorem on the operators on $L^2(\agb,\mu_o)$ defined by $\Gb$
invariant vector fields on $\Ab$.

\begin{theorem} Let $X$ be a $\Gb$ invariant vector field on $\Ab$.
The operator 
\be
P(X) \:= i(X + {1\over2}\div X)
\ee
with domain $\Cyl^1(\agb)$ is essentially self-adjoint on $L^2(\agb,
mu_o)$.  Suppose $Y$ is another $\Gb$ invariant vector field on
$\agb$; then, $[X,Y]$ is also a $\Gb$ invariant vector field on $\agb$
and
\be
P([X,Y])\ =\ i[P(X),P(Y)].
\ee
on $\Cyl^2(\agb)$.

\end{theorem}
\bigskip\goodbreak  
\subsection{Characterization of vector fields on $\agb$}

In the previous subsection we showed that the $\Gb$-invariant vector
fields on $\Ab$ have interesting properties. It is therefore of
considerable interest to have control on the structure of such vector
fields. Can one construct them explicitly? What is the available
freedom? To answer such questions, we will now obtain a complete
characterization of the $\Gb$-invariant vector fields on $\Ab$
in the case when $G$ is assumed to be semi-simple.

Fix a graph $\g_0$. To construct a $\Gb$-invariant vector field $X
=(X_\g)_{\g\ge\g_0}$ on $\Ab$, we have, first of all, to specify a
$\G_{\g_0}$-invariant vector field on $\A_{\g_0}$. We want to analyze
the freedom available in extending this vector field to $\A_\g$ for all
$\g\ge \g_0$. Now the edges of any $\g\ge \g_0$ can be ordered in such
a way that:
\begin{enumerate}
\item the first $n$ edges, ${e}_1,...,e_{n}$, are contained in $\g_0$,
for an appropriate $n$;
\item the next $m-n$ edges, ${e}_{n+1},...,e_{m}$, begin at (i.e., have
one of their vertices) on $\g_0$, for some $m$; and
\item the remaining edges, say, $e_{m+1}, ..., e_{k}$, do not intersect 
$\g_0$ at all.
\end{enumerate}
Hence, we can decompose $\A_\g$ as:
\be
{\A_{\g}}= \A_{\g_1}\times G^{m-n}\times G^{k-m}\ ,
\ee
where $\g_1\geq\g_0$ is the graph formed by the first $n$ edges.
Given $X_{\g_0}$, the projection $\A_{\g_1} \rightarrow \A_{\g_{0}}$
determines --via consistency conditions-- the vector $X_{\g_1}$ modulo
a vector field tangent to the fibres of the projection. But the fibres
are contained in the orbits of the symmetry group $\G_{\g_1}$. Hence,
from the point of view of operators on cylindrical functions on
$\agb$, this ambiguity is irrelevant; there is no essential freedom in
extending the vector field from $\A_{\g_0}$ to $\A_{\g_1}$. Next,
consider the last $k-m$ components of $X_{\g}$ corresponding to the
last set of edges. Since both vertices of these edges lie outside
$\g_0$, the corresponding vector fields on $G$ have to be both right
and left invariant. Since we have assumed that the gauge group is
semi-simple, this implies that they must vanish. Thus, the essential
freedom in extending $X_{\g_0}$ to $X_{\g}$ lies in the $m-n$
components of $X_\g$ associated with the second set of edges. Hence,
using the notation of Lemma \ref{3}, we can express $X_\g$ as:
\be
X_\g\ :=\ (X_{\g_0}, F([e_{n+1}]),..., F([e_{m}]),0,...,0)\ . 
\ee
Here, $F([e_i])$ with $i =n+1, ..., m$ is a function 
$$F([e_i])\ :\   \A_{\g_1}\ \rightarrow \ \Gamma(T(\A_{e_i})) $$
whose values are $\G_{v_i}$ invariant vector fields on $\A_{e_i}$,
where $v_i$ is the end of the edge $e_i$ which is not contained in
$\g_0$. The $\Gb$ invariance of $X$ implies that $F([e_i])$ should
have certain transformation properties under the action of the groups
$\G_v$ which act on the fibers over the vertices $v$ of $\g_0$.  Given
$a_v\in \G_v$ , we need: $F([e_i]) \circ a_v = (a_v)^{-1}_\star {F}
([e_i])$ if $v$ is the vertex of $e_i$ and $F([e_i])\circ a_v =
F([e_i])$ otherwise.

We can now summarize the information that is necessary and sufficient
to define a $\Gb$-invariant vector field $X$ on $\Ab$. First, we need
a graph $\g_0$. For each $x\in \g_0$, let $e_x$ be the set of germs of
edges (i.e., the data at $x$ that is necessary and sufficient to
specify edges) 
which do not overlap with any of the edges of $\g_0$ that pass through
$x$.  (Recall that edges are all analytic.) Let $\P_{\g_0}$ be the
sheaf of germs of transversal edges over the $\g_0$ divided by the
reparametrizations, i.e., set
\be
\P_{\g_0}\ \ :=\ \ \bigcup_{x\in\g_0}\P_x \ .
\ee
Next, given a point $x$ on $\g_0$ let $\g_x\ge \g_0$ be the graph
obtained by cutting the edge on which $x$ lies into two at $x$. (If
$x$ is a vertex of $\g_0$, then $\g_x =\g_0$.) Finally, choose a point
in the fiber of the underlying bundle $B(\S,G)$ over each point of
every edge on $\g_0$. Up to this freedom, the group $\G_x$ is then
identified with $G$. Then, the necessary and sufficient data for
constructing a $\Gb$-invariant vector field $X= (X_\g)_{\g\ge \g_0}$
(regarded as an operator on $\Gb$-invariant function on $\Ab$) is the
following
\begin{enumerate}
\item A $\G_{\g_0}$-invariant  vector field $X_{\g_0}$ on $\A_{\g_0}$;
\item For every graph $\g_1$ obtained from $\g_0$ simply by subdividing
the edges, a vector field $X_{\g_1}$ such that $p_{\g_0\g_1*}X_{\g_1} =
 X_{\g_0}$; 
\item A map from the set of germs of transversal edges $\P_{\g_0}$
into the Lie algebra  (of left invariant vector fields on $G$)
$LG$-valued functions on $\A_{\g_x}$, 
\be
F: \ \ \P_{\g_0} \rightarrow C^{n}(\A_{\g_x})\otimes \LG;
\ee
which has the following transformation properties with respect to the 
group $\G_{\g_x}$:
\be
F([e]_x)\circ a_v= \cases {F([e]_x), &if $v\not= x$\cr
                               a^{-1}F([e]_x)a, & if $v=x$\cr}
\ee
for every vertex $v$ of $\g_x$.
\end{enumerate}

 Then, given an edge $e$ intersecting
$\g$ at $x$ and the corresponding space $\A_{e}$, a value of $F([e])$
defines unambiguously a $\G_{v'}$ invariant vector field on $\A_e$,
$v'$ standing for the other end of $e$ (because the remaining freedom
of a gauge for $\A_e$ is covered by the action of $\G_{v'}$).

Thus, there is a rich variety of $\Gb$ invariant vector fields on
$\Ab$.  
The vector field is $\Gb$ invariant If the co-metric tensor and the
1-form are $\Gb$ invariant, so is the vector field.

\medskip

We will conclude this discussion of vector fields by pointing out
that, if one is interested only in the action of the vector fields on
cylindrical functions, apriori there appears to be some freedom in
one's choice of the initial definition itself. There are at least
three ways of modifying the definition we used.
 
\begin{enumerate}
\item First, we could have chosen another set of labels. Our definition
used graphs $\g\ge\g_0$ for some $\g_0$. Instead, we could have
labelled the vector fields by {\it any} cofinal $\Gra(\S)$.
However, it is not clear if all our results would go through in this
more general setting. In particular, it is not obvious that the vector
fields would then form a Lie algebra.
\item
Another possibility is to use the same labels ($\g\ge\g_0$ for some
$\g_0$) but to weaken the consistency conditions slightly. Since we
only want to act these vector fields $(X_\g)_{\g\ge\g_0}$ on functions
which are $\Gb$ invariant, it would suffice to require only that the
consistency conditions are satisfied ``modulo the gauge directions''.
That is, one might require only that each $X_\g$ is $\G_\g$ is
invariant and $p_{\g\g'\star}X_{\g'} = X_\g$ if $\g'\ge \g$, {\it both
modulo the directions tangent to the fibers of the group $\G_\g$}.
However, then it is no longer clear that the notion of divergence of
$X$ is well-defined. Further work is needed.
\item
Finally, throughout this paper, we have considered projective families
labelled by graphs. Alternatively one can also consider projective
families labelled by subgroups of the group of equivalence classes of
closed, based loops in $\S$, where two loops are equivalent if the
holonomy of any connection around them, evaluated at the base point,
is the same. (See \cite{MM,AL2}.) One can define $\Gb$-invariant
vector fields in this setting as well and the resulting momentum
operators on $L^2(\agb, \mu_o)$ are essentially the same as those
introduced here. However, the proofs are more complicated since they 
essentially involve decomposing loops in to graphs used here.
\end{enumerate}

\section{Laplacians, heat equations and heat kernel
measures on $\agb$ associated with edge-metrics.}

In the last two sections, we saw that, although $\Ab$ and $\agb$
initially arise only as compact topological spaces, using graphs on
$\S$ and the geometry of the Lie group $G$ one can introduce on them,
quite successfully, structures normally associated with manifolds.
Therefore, a natural question now arises: can one exploit the
invariant Riemannian geometry on $G$ to define on $\agb$ new
structures. In this section, we will show that the answer is in the 
affirmative.

\subsection{A Laplace operator.}

Let us fix an (left and right) invariant metric tensor $k$ on $G$.
The obvious strategy --which, e.g., successfully led us to the Haar
volume form in section 4.2-- would be to use the fact that $\A_\g$ can
be identified with $G^E$, to endow it with the product metric
${k}'_\g$ and let ${\Delta}'_\g$ be the associated Laplacian operator.
(Here, as before, $E$ is the number of edges in the graph $\g$.)
Unfortunately, this strategy fails: the resulting family of operators
$({\Delta}'_\g)_{\g\in\Gra(\S)}$ fails to be self consistent. (This is
why we have used the prime in $k'$ and ${\Delta}'$.) This is a
good illustration of the subtlety of the consistency conditions and
brings out the non-triviality of the fact that the families that led
us to forms, vector fields and the Haar measure turned out to be
consistent.
 
The ``minimal'' modification that leads to a Laplacian requires an
additional ingredient: a metric on the space of edges on $\S$. An {\it
edge-metric on $\S$}, will be a map which assigns to each edge (i.e.,
finite, analytic curve) $e$ in $\S$ a non-negative number, $l(e)$
which is independent of the orientation of $e$ and additive, i.e.
satisfies:
\begin{equation}
l(e^{-1})\ =\ l(e)\ \ \ {\rm and}\ \ \ l(e_1\circ e_2)= l(e_1)
+l(e_2)\ .
\end{equation}

\medskip\noindent
$l$ can be thought of as a generalized ``length'' function on the
space of edges.  The technique of using such  ``an additive weight''
was suggested by certain methods employed by Kondracki and Klimek
\cite{KK} in the context of 2-dimensional Yang-Mills theory.

It is not difficult to construct edge-metrics explicitly.  Two simple
examples of such constructions are:
\begin{enumerate}
\item Introduce a Riemannian metric $g$ on $\S$ and let  
$l(e)$ be the length of $e$.  
\item Fix a collection $s$ of analytic surfaces in $\S$ and define $l(e)$
to be the number of isolated points of intersection between $e$ and
$s$.
\end{enumerate}
\bigskip

Given an edge-metric $l$, for each  graph $\g$ we define on 
$\A_\g(=G^E)$ the following ``weighted'' Laplacian,
\begin{equation}
\Delta_{\g,(l)}\ :=\ l(e_1)\Delta_{e_1}+...+ l(e_E)\Delta_{e_E}\ ,
\end{equation}
where $\Delta_{e_i}$ is a an operator which applied to a function
$f_\g(g_{e_1},...,g_{e_E})$ acts (only) on the $G$-variable $g_{e_i}$
as the Laplacian of the metric $k$ on $G$. It is not obvious that 
this $\Delta_{\g,(l)}$ is well-defined since the isomorphism between
$\A_\g$ by $G^E$ used in the above construction is not unique.
However, two such isomorphisms are in essence related by an element of
$\G_\g$ and since $\Delta$ is left and right invariant on $G$,
$\Delta_{\g,(l)}$ is well-defined; it is insensitive to the ambiguity
in the choice of $(g_{e_1},...,g_{e_E})$ that label the points of
$\A_\g$. Furthermore, we have:

\begin{theorem}{\rm :} \label{delta} 

\noindent $(i)$ The family of operators $(\Delta_{\g,(l)},  
C^2(\A_\g))_{\g\in \Gra(\S)}$ is self consistent (in the sense of 
(\ref{01}) and  (\ref{02}));

\noindent $(ii)$
The operator $(\Delta_{(l)},\Cyl^2(\overline{\A}))$ defined in the 
Hilbert space $L^2(\Ab, \mu_o)$, where $\mu_o$ is the Haar volume 
form on $\Ab$, by
\begin{equation}
\Delta_{(l)}\ ([f_\g]_\sim):=[\Delta_{\g,(l)}\ (f_\g)]_\sim 
\end{equation}
is essentially self-adjoint; 

\noindent $(iii)$
the operator $\Delta_{(l)}$ is $\Gb$ invariant and thus defines an 
essentially self-adjoint operator $(\Delta_{(l)},\Cyl^2(\agb))$ in 
$L^2(\agb,\mu'_o)$, where $\mu'_o$ is the push-forward of $\mu_o$ to
$\agb$. 
\end{theorem}

{\it Proof}{\rm:} Note first that, for every $\g'\ge\g$, the
projection $p_{\g\g'}:\A_{\g'}\rightarrow \A_\g$ can be written as a
composition of projections, 
\be
p_{\g\g'}\ =\ p_{\g\g_1}\circ...\circ p_{\g_{n}\g'}
\ee
each of the terms being of a one of the following three kinds:
\be
(g_1,...,g_k, ...)\ \mapsto\ (g_1,...,(g_k)^{-1}, ...),\ \ \ 
(g_1,...,g_{k-1},g_k, ...)\ \mapsto\ (g_1,...,g_{k-1}, ...),
\ee 
\be
(g_1,...,g_{k-1},g_k)\ \mapsto\ (g_1,...,g_{k-1}g_k).
\ee
The operators $\Delta_{l(e),\g}$ are automatically consistent with
projections of the second class above; no conditions need to be
imposed on $l(e_i)$.  However, to be consistent with a projection of
the third kind, the numbers $l(e)$ have to satisfy the following
necessary and sufficient condition. Let $f$ be a function on $G$.
Whenever we divide an edge $e$ into $e=e_2\circ e_1$, then
\be
(l(e_1)\Delta_{e_1} \ +\ l(e_2)\Delta_{e_2})f(g_2, g_1)\ =
\ (l(e)\Delta f)(g_2 g_1)\ ,
\ee
where $f(g_2, g_1) = f(g_2 g_1)$. A short calculation shows that the
necessary condition for this to hold is precisely our second
restriction on $l(e)$: $l(e)=l(e_1)+l(e_2)$.  Similarly, our first
restriction $l(e^{-1}) =l(e)$ is necessary and sufficient to ensure
consistency with respect to the projections of the first type. This
establishes part $(i)$ of the theorem.

Part $(ii)$ follows easily from Lemma 1 and from the essential
self--adjointness of the operators $(\Delta^{(l)}_\g, C^2(\A_\g))$
defined in $L^2(\A_\g, (\mu_H)^E)$. Part $(iii)$ is essentially
obvious.

\bigskip

{\it Remark:} If $l(e)$ is non-zero for every non-trivial edge --as is
the case for the our first example of $l(e)$-- we can introduce on
each $\A_\g=G^E$ a block diagonal metric tensor $k_\g\ :=\ ({1\over
l(e_1)}k,...,{1\over l(e_E)}k)$. The operator $\Delta^{(l)}_\g$ is
just the Laplacian on $\A_\g$ defined by $k_\g$.

\subsection{The heat equation and the associated volume forms.}

Given an edge metric $l(e)$, we have a Laplacian on $\Ab$. It is now
natural to use these Laplacians to formulate heat equations on $\Ab$,
to find the corresponding heat-kernels and to construct the associated
heat-kernel measures on $\Ab$. These would be natural analogs of the
usual Gaussian measures on topological vector spaces.

As usual, heat equations will involve a parameter $t$ with $0<t$ and
a Laplacian $\Delta_{(l)}$ on $\Ab$. A 1-paramter family of functions
$f_t\in\Cyl^2(\Ab)$ will be said to be a solution to the heat equation
on $\Ab$ if it satisfies:
\be
{d\over dt}f_t\ =\ \Delta_{(l)}\ f_t  \ .
\ee  

Our main interest lies in defining a heat kernel for this equation.
Recall first that the heat kernel $\rho_t$ on $G$ is the solution to
the heat equation on $G$ satisfying the initial condition $\rho_{t=0}
= \delta(g, I_G)$, where $I_G$ is the identity element of $G$ and
$\delta$ is the Dirac distribution. As is well-known \cite{S}, for
each $t$, $\rho_t$ is a positive, smooth function on $G$. The next
step is to use $\rho_t$ to find a heat kernel $\rho_{\g,t}$ on
$(\A_{\g}, \Delta_{\g, (l)})$. This $\rho_{\g,t}$ is a function on
$\A_\g \times \A_\g$. Using a coordinatization of $\A_\g$ induced by a
group valued chart $\A_\g = G^E$, we can express $\rho_{\g,t}$ as:
\be \label{k1}
\rho_{\g, t}(A_\g,B_\g)\ =\ \rho_{s_1}(b_1^{-1}a_1)\ \ 
...\ \ \rho_{s_E}(b_E^{-1}a_E),
\ee
where, $A_\g=(a_1,...,a_E)$, $B_\g=(b_1,...,b_E)$ and where $s_i$ with
$i= 1, ..., E$ are positive numbers given by $s_i= l(e_i)t$.  (Note
that, since the heat kernel $\rho_t$ on $G$ is $\Ad(G)$ invariant, the
above formula is independent of the choice of coordinates.)  The
family $(\rho_{\g,t})_{\g\in\Gra(\S)}$ does not define a cylindrical
function on $\Ab$.  However, we can consider the convolution of the
heat kernel $\rho_{\g, t}$ with a function $f_\g\in C^0(\A_\g)$. In
the coordinatization used above, this reads:
\be \label{k2}
(\rho_{t,\g}\star f_\g)(A_\g)\ =\ 
\int_{A_\g}\rho_{t,\g}(A_\g,B_\g) f(B_\g)\mu_H(B_\g)\ ,
\ee   
where $\mu_H$ is the Haar measure on $\A_\g$. Now, the key point is
that these convolutions do satisfy the consistency conditions:
\be  
f_{\g_1}\ \sim\ f_{\g_2}\quad \Rightarrow\quad \rho_{t,\g_1}\star 
f_{\g_1}\ \sim\  \rho_{t,\g_1}\star f_{\g_1}\ .
\ee
Therefore, we {\it can} define a convolution map $\rho_t \star$ on
$\Ab$.

\begin{theorem}{\rm :}  \label{heat}
Let $\Delta_{(l)}=(\Delta_{\g, (l)})_{\g, \g'\in\Gra(\S)}$ be the
Laplace operator on $\Ab$ given by the edge-metric $l$ and let
$(\rho_{\g, t})_{\g\in \Gra(\S)}$ be the corresponding family of
heat kernels, with $t\ge 0$. Then,

\noindent(i)there exists a linear map $\rho_t\star\ :\ C^0(\Ab)\ 
\rightarrow\ C^0(\Ab)$ defined by 
\be \label{convolution}
\rho_t\star f\ = \ [\rho_{t,\g}\star f_\g]_\sim\ \ ,
\ee
which is continuous with respect to the sup-norm; 
\noindent(ii) If $f\in\Cyl^2(\Ab)$ then 
\be
f_t\ :=\ \rho_t\star f
\ee
solves the heat equation with the initial value $f_{t=0} =f$;

\noindent(iii) The map $\rho_t\star$  carries $C^0(\agb)$ 
(the $\Gb$ invariant elements of $C^0(\Ab)$) into 
$C^0(\agb)$. 
\end{theorem}

{\it Proof}{\rm :} To establish $(i)$, it is sufficient to prove the
 right hand side of (\ref{convolution}) is well defined. This would
imply that $\rho_t^\star$ is well-defined on $C^0(\Ab)$. Continuity
follows from the explicit formula for the convolution.

Let $f=[f_{\g_1}]_\sim = [f_{\g_2}]_\sim\in\Cyl^2(\Ab)$. Choose any
$\g'\ge\g_1,\g_2$. Using the consistency conditions satisfied by the
family $(\Delta_{\g, (l)})_{\g,\g'\in\Gra(\S)}$ which are guaranteed
by Theorem \ref{delta}, and setting $i=1,2$, we have:
\be
{d\over dt}\ p_{\g_i\g'}^\star\ (\rho_{t,\g_i}\star f_{\g_i})=
p_{\g_i\g'}^\star\ (\Delta_{\g_i,(l)}\ \rho_{\g_i, t}\star f_{\g_i}) =
\Delta_{\g', (l)}\ p_{\g_i\g'}^\star\ (\rho_{\g_i, t}\star f_{\g_i})\ ,
\ee 
on $\A_{\g'}$. Thus, we conclude that the pull-backs of the two
convolutions satisfy the heat equation on $\A_{\g'}$ with initial
values $p_{\g_1\g'}^\star\ f_{\g_1}$ and $p_{\g_2\g'}^\star f_{\g_2}$
respectively. However, since $f_{\g_1}\sim f_{\g_2}$, the two initial
values are equal. Hence, the two pulled back convolutions on
$\A_{\g'}$ are equal. Finally, since
\be
p_{\g_1\g'}^\star\ (\rho_{t,\g_1}\star f_{\g_1})\ =\ 
\rho_{t,\g'}\star p_{\g_1\g'}^\star f_{\g_1}\ =\ p_{\g_2\g'}^\star\ 
\rho_{t,\g_2}\star f_{\g_1} \ ,
\ee
we conclude that the right side of (\ref{convolution}) is
well-defined.

Part $(ii)$ is now obvious and part $(iii)$ follows from the $\Gb$
symmetry of $\Delta^{(l)}$.

$\Box$

\bigskip

We can now define the heat-kernel volume forms on $\Ab$ associated to
the Laplace operator $\Delta_{l}$. These are the natural analogs of
the Gaussian measures on Hilbert spaces, which can also be obtained as
a projective limit of Gaussian volume forms on finite dimensional
subspaces. Now, on a finite dimensional Hilbert space, the natural
Gaussian functions $g(\vec{x})$ are given by the heat kernel,
$g(\vec{x}) = \rho_{t}(o,\vec{x})$, and the Gaussian volume form is
just the product of the translation invariant volume form by
$g(\vec{x})$.  The heat-kernels forms on $\A_{\g}$ will be obtained
similarly by multiplying the Haar volume form by the heat kernel.
Unlike a Hilbert space, however, $\A_{\g}$ does not have a preferred
origin. Let us therefore first fix a point $\bar{A}_0=(A_{0\g})_{\g\in
\Gra(\S)}\in\Ab$.  Then, on each $\A_\g$, we define a volume form
$\nu_\g$ which, when evaluated at the point $A_\g\in \A_\g$, is given
by
\be \label{gaussian}
\nu_{t,\g}(A_\g)\ :=\ \rho_{t,\g}(A_{0\g}, A_\g)\mu^H_\g(A_\g)
\ee 
where $\mu^H_\g(A_\g)$ denotes the Haar form on $\A_\g$ at $A_\g$.
We have
\begin{theorem}{\rm:} 
\noindent (i)
The family $(\nu_{t,\g})_{\g\in\Gra(\S)}$ of volume forms defined by
(\ref{gaussian}) is consistent;

\noindent (ii) the measure, $\nu_t\ :=\ (\nu_{t,\g})_{\g\in\Gra(\S)}$, 
defined on $\Ab$ by the volume form is faithful iff for every non-zero
edge $e$ in $\S$, $l(e)>0$.
\end{theorem}

The proof follows from Theorem \ref{heat}, and from the nonvanishing
of the heat kernel on $G$.
\medskip

Thus, to define a heat-kernel volume form $\nu_t$ we have used the
following input: an invariant metric tensor on $G$, an edge-metric $l$,
a point $\bar{A}_0\in\Ab$ and a `time' parameter $t>0$. In terms of
the heat kernel of $\Delta^{(l)}$, the integral of a function
$C^0(\Ab)\ni f=[f_\g]_\sim$ with respect to $\nu_t$
is given by the formula
\be \label{gmeasure}
\int_{\Ab}f\ =\ (\rho_{t}\star f)(\bar{A}_0) ,
\ee
which is completely analogous to the standard formula for integrals
of functions on Hibert spaces with respect to the Gaussian measures.

Finally, let us consider the quotient $\agb\ =\ \Ab/\Gb $.  According
to Theorem (\ref{heat}), the heat kernel $\rho_t$ naturally projects
to this quotient and hence we have a well-defined Gaussian measure
$(\ref{gmeasure})$ on $\agb$. To have a $\Gb$ invariant volume form on
$\Ab$ corresponding to this measure, we just average the Gaussian
volume form on $\Ab$, following the procedure of section 5.1,
\be
\overline {\nu_t}\ =\ \int_\Gb (R_{\bar g\star}\ \nu_t) d{\bar g}\ ,
\ee
where $d\bar{g}$, as before, is the Haar form on $\Gb$.  The resulting
volume form $\overline {\nu_t}$ shall be referred to as a Gausian
volume form on $\agb$. Finally, on $\agb$, there is a natural origin
$[\bar{A}_0]$ whence the freedom is the choice of $\bar{A}_0$ can be
eliminated.  To see this, recall first that while constructing a heat
kernel measure on a group, one picks the identity element as the
fiducial point.  The obvious analog in the present case would be to
choose $\bar{A}_0\in\Ab$ such that the parallel transport given by
$\bar{A}_0$ (see Proposition \ref{Hol}) along any closed loop in $\S$
is the identity. This does not single out $\bar{A}_0$ uniquely in
$\Ab$. However, all points $A_0\in \Ab$ with this property project
down to a single point in the quotient $\agb=\Ab/\Gb$.

\bigskip

The techniques introduced in this section can be extended in several
directions. In the next subsection, we present one such extension,
where heat kernel measures are defined on the projective limit of the
family associated with {\it non-compact} structure groups $G$.
Another extension has been carried out in \cite{ALMMT3} where
diffeomorphism invariant heat kernels are introduced using, in place
of the induced Haar measure on $\Ab$, the diffeomorphism invariant
measures introduced by Baez \cite{B1,B2}.
 
\subsection{Non-compact structure groups.} 

We now wish to relax the assumption that the structure group be
compact and let $G$ be any connected Lie group.  We will see that we
{\it can} extend the construction of the induced Haar measure as well
as the heat kernel measures to this case, although there are now
certain important subtleties.

To begin with, the construction of the space $\agb$ given in \cite{AI}
does not go through, since the Wilson loop functions are now unboun[216zded
and one can not endow them with the structure of a $\C^\star$-algebra in
a straightforward fashion. However, following Mour\~ao and Marolf
\cite{MM}, we can still introduce the projective family $(\A_\g,
\G_\g,p_{\g\g'})_{\g, \g'\Gra(\S)}$ of analytic manifolds of
section 3. As in section 3, the members $\A_{\g}$ of this family have
the manifold structure of $G^{E}$ (where $E$ is the number of edges in
$\g$) and are therefore no longer compact. Nonetheless, the projective
limit is well-defined and the notion of self-consistent families of
measures $(\mu_{\g})_{\g\in\Gra(\S)}$ is still valid. Each such family
enables us to integrate cylindrical functions. However, since the
projective limit is no longer compact, the proof \cite{AL2} that each
self-consistent family defines a regular Borel measure $\mu$ on the
limit does not go through. Thus, in general, the family only provides
us with cylindrical measures on the projective limit. Nonetheless, the
construction of these measures is non-trivial since the consistency
conditions have to be solved and the result is both structurally
interesting and practically useful.

Let us first construct some natural measures on $\Ab$. In the previous
constructions, we began with the Haar measure on $G$. However, since
$G$ is now non-compact, its Haar measure can not be normalized and we
need to use another measure as our starting point. Fix any probability
measure $d\mu_1$ on $G$ which is $\Ad(G)$ invariant and has the
following two properties:
\be
\int_{G^2} f(gh)d\mu_1(g)d\mu_1(h) \ =\ \int_G f(g)d\mu_1(g),\ \ 
{\rm and} \ \ d\mu_1(g)\ =\ d\mu_1(g^{-1})
\ee
for every $f\in C^0_0(G)$ (the space of continuous functions on $G$
with compact support). Then, the procedure introduced in \cite{AL1}
provides us with a family of measures $(\mu_\g)_{\g\in\Gra(\S)}$, each
$\mu_{\g}$ being the product measure on $\A_\g=G^E$. In spite of the
fact that $\A_{\g}$ are non-compact, results of \cite{AL1} required to
ensure the self consistency of the family are still applicable. We
thus have an integration theory on $\Ab$. The resulting cylindrical
measure is again faithful and invariant under the induced action of
Diff$(\S)$.

Another possibility is to use the Baez construction \cite{B1} which
leads to a diffeomorphism invariant integration on $\agb$ from almost any
measure on $G$. The resulting cylindrical measures, however, fail to
be faithful.

A third possibility is to repeat the procedure of gluing measures on
$\A_{\g}$ starting, however, from an appropriately generalized heat
kernel measure on $G$. Given a measure $\mu$ on $G$, let us define an
integral kernel, $\rho$ via:
\be
(\rho_t\star f)(g)\ :=\ \int_G f(gh^{-1})d\mu(h) \ . 
\ee
A 1-parameter family of measures $\mu_t$ on $G$ will be said to
constitute a {\it generalized heat kernel measure} if the resulting
integral kernels $\rho_t$ satisfy the following conditions:
\be
\rho_r\star \rho_s\star f\ =\ \rho_{r+s}\star f, \ \ 
\rho_t\star R_{g\star}f\ =\ R_{g\star}\rho_t\star f,\ \  
\rho_t\star I_{\star}f\ =\ I_{\star}\rho_t\star f \ ,
\ee
where $R$ and $I$ denote, respectively, the right action of $G$ in $G$
and the inversion map $g\mapsto g^{-1}$, and where $f\in C^0_0(G)$
%
%

Given an edge-metric $l$ on $\S$, we can now repeat the construction
of the previous subsection and obtain a self consistent family of
measures $(\mu_{\g})_{\g\in{\Gra(\S)}}$. Thus,
for a graph $\g$, define the following ``generalized heat
evolution''
\be \label{he}
(\rho_{t,\g}\star f_\g)(A_\g)\ =\ \int_{\A_\g} f_\g(a_1 b_1^{-1}, 
...,a_E b_E^{-1})\  d\mu_{s_1}(b_1)...d\mu_{s_E}(b_E)
\ee
where $s_i = l(e_i)t$ and where we have set $A_\g=(a_1,...,a_E)$,
$B_\g=(b_1,...,b_E)$ using an identification $\A_\g=G^E$. It is then
straightforward to establish the following result.
\begin{proposition}{\rm :} The family of heat evolutions (\ref{he})
in self consistent, i.e., if $f$ is a cylindrical function,
$f=[f_{\g_1}]_\sim= [f_{\g_2}]_\sim$, then
\be  
 [\rho_{t,\g_1}\star f_{\g_1}]_\sim\ =\  [\rho_{t,\g_2}\star 
 f_{\g_2}]_\sim.
\ee
\end{proposition}

Using the resulting heat evolution for $\Ab$ we can just define
the corresponding heat kernel integral to be
\be
\int_\Ab  =\ \rho_t\star f(\A_0)
\ee
as before. If the group $G$ is compact, this formula defines a regular
Borel measure on $\Ab$. In the general case, we only have a
cylindrical measure. A case of special interest is when $G$ is the
complexification of a compact, connected Lie group. It is discussed
in detail in \cite{ALMMT3}.

\bigskip\goodbreak

\section{Natural geometry on $\agb$.}

Let us begin by spelling out what we mean by ``natural structures'' on
$\Ab$ and on $\agb$.

A natural structure on $\ab$ will be taken to be a structure which is
invariant under the induced action of the automorphism group of the
underlying bundle $B(\S,G)$. (Note that {\it all} automorphisms are
now included; not just the vertical ones.) In the context of the
quotient, $\agb$, a natural structure will have to be invariant under
the induced action of the group Diff($\S$) of diffeomorphisms on $\S$.

The action of Diff($\S$) on $\agb$ may be viewed as follows. Let
$\Phi\in \Diff(\S)$.  Given a graph $\g=\{e_1,...,e_E\}$ choose any
orientation of the edges and consider the graph $\Phi(\g) =
\{\Phi(e_1),...,\Phi(e_E)\}$ with the corresponding orientation. Pick
any trivialization of $B(\S,G)$ over the vertices of $\g$ and
$\Phi(\g)$ and define
\be
\Phi_\g\ :\ \A_\g/\G_\g\ \rightarrow\ \A_{\Phi(\g)}/\G_{\Phi(\g)},\ \ \
\Phi_\g\ :=\   \Lambda_{\Phi(\g)}^{-1} \circ \Lambda_\g,
\ee
where $\Lambda_\g$ is the group valued chart defined in (\ref{diff}).
It is easy to see, that the family of maps $\Phi_\g$ defines uniquely
a map
\be
\overline{\Phi}\ :\ \agb\ \rightarrow\ \agb.
\ee
This is the action of $\Diff(\S)$ on $\agb$.  
\bigskip

In this section we will introduce a natural contravariant, second
rank, symmetric tensor field $k_o$ on $\agb$ and, using it, define a
natural second order differential operator. In the conventional
approaches, by contrast, the introduction of such Laplace-type
operators on (completions of) $\ag$ always involve, to our knowledge,
the use of a background metric on $\S$ (see, e.g. \cite{AM}).

\goodbreak
\subsection{Co-metric tensors.}  

Let us suppose that we are given, at each point $A_\g$ of $\A_\g$, a
real, bilinear, symmetric, contravariant tensor $k_\g$:
\be
k_\g \ :\ \bigwedge{}_{A_\g}\times \bigwedge{}_{A_\g}\ \rightarrow
\ \RR,
\ee  
for all graphs $\g$, where $\bigwedge_{A_\g}$ denotes the cotangent
space of $\A_\g$ at $A_\g$. Such a family, $(k_\g)_{\g\in L}$, will be
called a {\it co-meric tensor} on $\ab$ if it satisfies the following
consistency condition,
\be
k_{\g'}(p_{\g\g'}^*\o_\g\wedge p_{\g\g'}^*\nu_\g)\ =\ 
k_{\g}(\o_\g\wedge\nu_\g), 
\ee
for every $\o_\g,\nu_\g\in\bigwedge_{A_\g}(\A_\g)$, whenever $\g'\ge\g$.

\bigskip
\noindent {\it Example.} Recall from Remark 1 at the end of section 
6.1 that, given a path metric $l$, and an edge $e$, one can introduce
on $\A_e$ a contravariant metric $l(e)k_e$, where $k_e$ is the
contravariant metric induced on $\A_e$ by a fixed Killing form on $G$.
Then, given a graph $\g$, $\A_\g=\times_{e\in\g}\A_e$ is equipped with
the product contravariant metric
 $$k^{(l)}_\g\ :=\ \sum_{e\in\g} l(e)k_e.$$ 
It is not hard to check that $(k^{(l)}_\g)_{\g\in L}$ is a co-metric
$k^{(l)}$ on $\Ab$. Notice, that if the edge metric vanishes for some
edges $e_0$, the co-metric is degenerate but continues to be well
defined.
\bigskip     

Given a co-metric $k=(k_\g)_{\g\in L}$ and a differential 1-form $\o =
[\o_\g]_\sim \in \Omega^1(\Ab)$, we define for each $\o_\g$ the vector
field $k_\g(\o_\g)$ on $\A_\g$ determined by the condition that, for
any 1-form $\nu$ on $\A_\g$,
$$
\nu(k_\g(\o_\g))\ =\ k_\g(\nu, \o_\g)
$$ 
at every $A_\g\in\A_\g$. It is easy to see that the family of vector
fields $(k_{\g'}(\o_{\g'}))_{\g'\ge \g}$ $=: k(\o)$ defines a vector
field on $\Ab$ in the sense of section 4.  Let us suppose that we are
given a co-metric tensor $k$ and a 1-form $\o$ on $\ab$ both of which
are $\Gb$ invariant. Then so is the vector field $k(\o)$.  Hence, we
can apply the results of section 5 to obtain

\begin{proposition}{\rm :} \label{d*} Suppose  $k$ is a $\Gb$ 
invariant co-metric tensor on $\Ab$ and $\o\in\Omega^1(\agb)$. Then
the vector field $k(\o)$ is compatible with the natural Haar volume
form $\mu_o$ on $\ab$.
\end{proposition}
 
Hence, assuming the necessary differentiability, $k(\o)$ has a well
defined divergence with respect to the Haar volume form $\mu_o$ on
$\ab$, $\div_{\mu_o} k(\o) = [\div k_\g(\o_\g)]_\sim$.  Consequently,
with every $\Gb$ invariant co-metric on $\Ab$ we may associate a
second-order differential operator
\be \label{d*d}
\Delta^{(k)}\ :\ \Cyl^2(\agb)\ \mapsto \Cyl^0(\agb);\ \ \ 
\Delta^{(k)}\ f = \div_{\mu_o}[k(df)]
\ee
with a well-defined action on the space of $\Gb$ invariant cylindrical
functions. If $k_\g$ is non-degenerate, $\Delta^{(k)}$ is the
Laplacian it defines. (In particular, the operator defined by the
co-metric $k^{(l)}$ via (\ref{d*d}) is precisely the Laplacian
constructed from the edge-metric $l$ in section 6.)  Hence, for
simplicity of notation, we will refer to $\Delta^{(k)}$ as the {\it
Laplace operator corresponding to $k$} even in the degenerate case.

\goodbreak
\subsection{A natural co-metric tensor.} 

We will now show that $\agb$ admits a {\it natural} (non-trivial)
co-metric. Let us begin by fixing an invariant contravariant metric on
the gauge group $G$ (i.e., an invariant scalar product in the
cotangent space at each point of $G$). This is the only ``input'' that
is needed. The idea is to use this $k$ to assign a scalar product
$k_v$ to each vertex $v$ of a graph $\g$ and set $k_\g= \sum_v k_v$.

Given a vertex $v$, choose an orientation of $\g$ such that all the
edges at $v$ are outgoing.  Use a group valued chart $\Lambda_\g$ (see
(\ref{diff})) to define for each edge $e$ in $\g$ a map
\be
G\ \rightarrow\ \A_e\ \rightarrow\ \times_{e'\in\g}\A_{e'}.
\ee
Through the corresponding pull-back map every $\o\in\bigwedge_{A_\g}$
defines a 1-form $\o_e$ on $G$ for every edge $e$. Let $k_e$ and
$k_{ee'}$ be two bilinear forms in $\bigwedge_{A_\g}$ defined by
\be
k_e(\o,\nu)\ =\ k(\o_e, \nu_{e}),\ \ \ k_{ee'}(\o,\nu)\ 
=\ k(\o_e, \nu_{e'})\ .
\ee
(Note that $k_e$ is as in the example in the last subsection.)  Both,
$k_e$ and $k_{ee'}$ are insensitive to a change of a group valued
chart since the field $k$ on $G$ is left and right invariant.  Now,
for $k_v$ we set
\be \label{sum}
k_v\ :=\ {1\over 2} \sum_e k_e\ + {1\over 2} \sum_{ee'} k_{ee'}
\ee
where the first sum is carried over all the edges passing through $v$
and, in the second, $e,e'$ range through all the pairs of edges 
which form an analytic arc at $v$ (i.e, such that $e'\circ e^{-1}$ is
analytic at $v$). Finally, we define $k_\g$ simply by summing over 
all the vertices $v$ of $\g$.
\be \label{k}
k_\g\ :=\ \sum_v k_v \ .
\ee

We then have
\begin{proposition}{\rm :} \label{naturalk} The family $k_o = 
(k_\g)_{\g\in L}$ is a natural co-metric tensor on $\ab$; $k_o$
is $\Gb$ invariant, and defines a natural co-metric on $\agb$.
\end{proposition}

The proof consists just of checking the consistency conditions and
proceeds as the previous proofs of this property. Note that
consistency holds only because we have added the terms $k_{ee'}$,
assigned to each pair of edges which constitute a single edge of a
smaller graph. (Unfortunately, however, this is also the source of the
potential degeneracy of $k_o$.) The diffeomorphism invariance of $k_o$
is obvious.

\goodbreak
\subsection{A natural Laplacian}

We can now use the natural co-metric to define a natural Laplacian.
That is, using Propositions \ref{d*} and \ref{naturalk} and Eq.
(\ref{d*d}), we can assign to $k_o$ the operator $(\Delta{\!}^o \equiv
\Delta^{(k_o)},\ \Cyl^2(\agb))$. In this subsection, we will
write out its detailed expression and discuss some of its properties.

Let us fix a group valued chart on $\A_\g$. The operator
$\Delta{\!}^o_{\g}$ representing $\Delta{\!}^o$ in $C^2(\A_\g/\G_\g)$ is
given by the sum
\be\label{D1}
\Delta{\!}^o_{\g}\ =\ \sum_v \Delta_v
\ee
where each $\Delta_v$ will involve products of derivatives on the
copies $\A_e$ corresponding to edges incident at $v$. To calculate
$\Delta_v$, we will orient $\g$ such that the edges at $v$ are all
outgoing. Let $R_{i}$, $i=1,...,n$ be a basis of left invariant vector
fields on $G$, $\theta^i$ the dual basis and let
$k^{ij}=k(\theta^i,\theta^j)$ denote the components of $k$ in this
basis.  Let $e$ be an edge with vertex $v$. Denote by $R_{ei}$ the
vector field on $\A_e$ corresponding to $R_i$ via the group valued
chart.  Next, identify $R_{ei}$ with the vector field
$(0,...,0,R_{ei},0,...,0)$ on $\A_\g=\times_{e'\in\g}\A_{e'}$.  Then,
the expression of $\Delta_v$ reads
\be \label{D2}
\Delta_v\ =\ {1\over 2} \sum_e k^{ij} R_{ei}R_{ej}
+{1\over 2}\sum_{ee'} k^{ij} R_{ei}R_{e'j} \ ,
\ee
where the sums are as in the expression (\ref{sum}) of $k_v$.

The consistency of the family of operators $(\Delta_{\g}{\!}^o, \Cyl^2
(\A_\g/\G_\g))$ defined by (\ref{D1}) and (\ref{D2}) is assured by
Theorem \ref{div2}. However, one can also verify the consistency
directly from these two equations. Indeed, if a vertex $v$ belongs to
only two edges, say $e$ and $e'$, which form an analytic arc (and are
oriented to be outgoing), then
\be
R_{ei}f_\g\ =\ -R_{e'i}f_\g
\ee
so that $\Delta_v=0$ at this vertex $v$. This ensures
\be \label{c}
p_{\g\g'}{}_\star \Delta{\!}^o_{\g'}\ =\ \Delta{\!}^o_\g,
\ee
when $\g'$ is obtained from $\g$ by splitting an edge. On the other
hand, if $\g'$ is constructed by adding an extra edge, say $e''$, to
$\g$ then the projection kills every term containing $R_{e''i}$. The
remaining terms coincide with those of $\Delta_\g$. Furthermore, in
this argument, we need not restrict ourselves to $\G_\g$ invariant
functions.  This shows that the vector field $k_o(df)$ is compatible
with the natural volume form $\mu_o$ on $\Ab$ for {\it every}
$f\in\Cyl^2(\ab)$. Hence, the operator
\be \label{Dext}
\Delta{\!}^o\ :\ \Cyl^2(\ab)\ \mapsto\ \Cyl(\ab); \ \ \Delta{\!}^o f =
\div_{\mu_o}[k_o(df)]
\ee
is also well defined. 

As in the Sec 6, the Laplace operator $\Delta{\!}^o$ can be used to
define a semi-group of transformations $\rho_{t\star}:\Cyl(\ab)
\rightarrow \Cyl(\ab) $ such that $f_t:=\rho_{t\star} f$ solves the 
corresponding heat equation.  In this case, $\rho_{t\star}$ is given
by the family $(\rho_\g)_{\g\in L}$ of certain generalized heat
kernels. The family and the transformations $\rho_{t\star}$ coincide
with those introduced in \cite{ALMMT3}. In fact the constructions of
this subsection were motivated directly by the results of
\cite{ALMMT3}.
\bigskip

We will conclude with two remarks. 
\begin{enumerate}
\item
If a vector field $X$ on $\ab$ is compatible with a volume form $\mu$,
then so is $Y= hX$ for any $h\in\Cyl^1(\ab)$ (and $\div(Y)=h\div(X)
+X(h)$). From this and from the existence of the operator
(\ref{Dext}), we have:
\begin{proposition}{\rm :}
For every (differentiable) 1-form $\o\in\Omega^1(\ab)$, the vector
field $k_o(\o)$ is compatible with the Haar volume form $\mu_o$.
\end{proposition}
\noindent The map
$$ d^\star\ :\ \Omega^1(\ab)\ \mapsto\ \Omega^0(\ab),\ \ d^\star \o =
\div_{\mu_o} [k_o(\o)] 
$$ 
can be thought of as a co-derivative defined on $\ab$ by the geometry
$(k_o,\mu_o)$.  It is possible to extend $d^*$ to act on
$\Omega^n(\ab)$.

\item
Consider the special case when the structure group $G$ is $SU(2)$ and
$\S$ is an oriented 3-dimensional manifold. There exists on $\ab$ a
natural  {\it third order} (up to the squer root)
differential  operator $({q}, \Cyl^3(\Ab))$ 
defined by a consistent family of operators $(q_\g, C^3(\A_\g))_{\g\in
L}$ on $C^3(\A_\g)$. To obtain $q_\g$, we begin as before, by defining
operators $q_v$ associated with the vertices of $\g$. Given a vertex
$v$ of a graph $\g$ and a triplet of edges $(e,e',e'')$ concident on
this vertex, let $\epsilon(e,e',e'')$ be $0$ whenever their tangents
at $v$ are linearly dependent, and $\pm 1$ otherwise, depending on the
orientation of the ordered triplet. To the vertex $v$, let us assign
an operator acting on $C^3(\A_\g)$ as
\be
q_v\ =\ \textstyle{1\over (3!)^2}\ \ \mu_H(R_i,R_j,R_k) 
\displaystyle{\sum_{(e,e',e'')}} \  \epsilon(e,e, e'')\  
R_{ei}R_{e'j}R_{e''k}
\ee
where we use the same orientation, notation and group valued charts as
in (\ref{D2}), and where the summation ranges over all the ordered
triples $(e,e',e'')$ of edges incident at $v$. In terms of these
operators, we can now define $q_\g$. Set
\be
q_\g\ :=\ \sum_v |q_v|^{1\over 2}
\ee
where $v$ runs through all the vertices of $\g$ (at which three or
more vertices have to be incident to contribute). We then have
\begin{proposition}{\rm :} The family $(q_\g, C^3(\A_\g))_{\g\in L}$ of
operators is self consistent. The operator $({q}, \Cyl^3(\ab))$ is
natural, $\Gb$ invariant and defines a natural operator on
$\Cyl^3(\agb)$.
\end{proposition}
\noindent This operator is closely related to the total volume 
(of $\S$) operator of Riemannian quantum gravity \cite{RS,ALMMT2}.
The detailed derivation of this operator as well as the local area
and the analytic formulae for the Hamiltonian
operators will be discussed in a forthcoming paper \cite{AL4}.
\end{enumerate}

\section{Discussion}

In more conventional approaches to gauge theories, one begins with
$\ag$, introduces suitable Sobolev norms and completes it to obtain a
Hilbert-Riemann manifold (see, e.g., \cite{AM}). However, since the
Sobolev norms use a metric on the underlying manifold $\S$ the
resulting Hilbert-Riemann geometry fails to be invariant with respect
to induced action of Diff$(\S)$. Hence, to deal with diffeomorphism
invariant theories such as general relativity, a new approach is
needed, an approach in which the basic constructions are not tied to a
background structure on $\S$. Such a framework was introduced in
\cite{AI,AL1} and extended in [5,8] to a theory of diffeomorphism
invariant measures. Here, the basic assumption is that the algebra of
Wilson loop variables should be faithfully represented in the quantum
theory and it leads one directly to the completion $\agb$. Neither the
construction of $\agb$ nor the introduction of a large family of
measures on it requires a metric on $\S$. This is a key strength of
the approach. The physical concepts it leads to are also significantly
different from the perturbative treatments of the more conventional
approaches: loopy excitations and flux lines are brought to forefront
rather than wave-like excitations and notion of particles.  The
framework {\it has been} successful in dealing with Yang-Mills theory
in 2 space-time dimensions \cite{ALMMT1} without a significant new
input. This success may, however, be primarily due to the fact that in
2 dimensions, the Yang-Mills theory is invariant under all volume
preserving diffeomorphisms. In higher dimensions, it remains to be
seen whether the Yang-Mills theory admits interesting phases in which
the algebra of the Wilson loop operators is faithfully represented. If
it does, they would be captured on $\agb$, e.g., through Laplacians
and measures which depend on an edge-metric $l(e)$, which itself would
be constructed from a metric on $\S$. We expect, however, that the key
applications of the framework would be to diffeomorphism invariant
theories.

A central mathematical problem for such theories is that of developing
differential geometry on the quantum configuration space, again
without reference to a background structure on $\S$.  This task was
undertaken in the last four sections. In particular, we have shown
that, although $\agb$ initially arises only as a compact Hausdorff
topological space, because it can be recovered as a projective limit
of a family of {\it analytic manifolds}, using algebraic methods, one
can induce on it a rich geometric structure.  There are strong
indications that this structure will play a key role in
non-perturbative quantum gravity.  Specifically, it appears that
results of this paper can be used to give a rigorous meaning to
various operators that have played an important role in various
heuristic treatments \cite{AA3,RS}.

We will conclude by indicating how the results of this paper fit
in the general picture provided by the available literature on the
structure of $\agb$. 

As mentioned in the Introduction, $\agb$ first arose as the Gel'fand
spectrum of an Abelian $C^\star$ algebra $\HAbar$ constructed from the
so-called Wilson loop functions on the space $\ag$ of smooth
connections modulo gauge transformations \cite{AI}. A complete and
useful characterization of this space \cite{AL1} can be summarized as
follows. Fix a base point $\xz$ on the underlying manifold $\S$ and
consider piecewise analytic, closed loops beginning and ending at
$\xz$. Consider two loops as equivalent if the holonomies of any
smooth connection around them, evaluated at $\xz$, are equal. Call
each equivalence class a {\it hoop} (a holonomic-loop). The space
$\HG$ of hoops has the structure of a group, which is called the {\it
hoop group}.  (It turns out that in the piecewise analytic case, $\HG$
is largely independent of the structure group $G$. More precisely,
there are only two hoop groups; one for the case when $G$ is Abelian
and the other when it is non-Abelian.) The characterization of $\agb$
can now be stated quite simply, in purely algebraic terms: $\agb$ is
the space of all homomorphisms from $\HG$ to $G$. Using subgroups of
$\HG$ which are generated by a finite number of independent hoops, one
can then introduce the notion of cylindrical functions on $\agb$. (The
$C^\star$-algebra of all continuous cylindrical functions turns out to
be isomorphic with the holonomy $C^\star$-algebra with which we
began.)  Using these functions, one can define cylindrical
measures. The Haar measure on $G$ provides a natural cylindrical
measure, which can be then shown to be a regular Borel measure on
$\agb$ \cite{AL1}.

Marolf and Mour\~ao \cite{MM} obtained another characterization of
$\agb$: using various techniques employed in \cite{AL1}, they
introduced a projective family of compact Hausdorff spaces and showed
that its projective limit is naturally isomorphic to $\agb$.  This
result influenced the further developments significantly.  Indeed, as
we saw in this paper, it is a projective limit picture of $\agb$ that
naturally leads to differential geometry.  The label set of the family
they used is, however, different from the one we used in this paper:
it is the set of all finitely generated subgroups of the hoop group.
At a sufficiently general level, the two families are equivalent, the
subgroups of the hoop group being recovered in the second picture as
the fundamental groups of graphs.  For a discussion of measures and
integration theory, the family labelled by subgroups of the hoop group
is just as convenient as the one we used. Indeed, it was employed
successfully to investigate the support of the measure $\mu_0$ in
\cite{MM} and to considerably simplify the proofs of the two
characterization of $\agb$, discussed above, in \cite {AL2}.  For
introducing differential geometry, however, this projective family
appears to be less convenient.

The shift from hoops to graphs was suggested and explored by Baez
\cite{B1,B2}. Independently, the graphs were introduced to analyze
the so-called ``strip-operators'' (which serve as gauge invariant
momentum operators) as a consistent family of vector fields in
\cite{Le2}. Baez also pointed out that, even while dealing with
gauge-invariant structures, it is technically more convenient to work
``upstairs'', in the full space of connections, rather than in the
space of gauge equivalent ones. Both these shifts of emphasis led to
key simplifications in the various constructions of this paper.

Baez's main motivation, however, came from integration theory. His
main result was two-folds: he discovered powerful theorems that
simplify the task of obtaining measures and, using them, obtained a
large class of diffeomorphism invariant measures on $\agb$.  In his
discussion, it turned out to be convenient to de-emphasize $\agb$
itself and focus instead on the space $\A$ of smooth connections. In
particular, he regarded measures on $\Ab$ primarily as defining
``generalized measures'' on $\A$. At first this change of focus
appears to simplify the matters considerably since one seems to be
dealing only with the familiar space of {\it smooth} connections. One
may thus be tempted to ignore $\agb$ altogether!

The impression, however, misleading for a number of reasons. First, as
Marolf and Mour\~ao \cite{MM} have shown, the support of the natural
measure $\mu_o$ is concentrated precisely on $\Ab$--$\A$; the space
$\A$ is contained in a set of zero $\mu_o$-measure. The situation is
likely to be the same for other interesting measures on $\agb$.
Thus, the extension from $\A$ to $\Ab$ is not an irrelevant
technicality.  Second, without recourse to $\Ab$, it is difficult to
have a control on just how general the class of ``generalized
measures'' on $\A$ is.  Is it perhaps ``too general'' to be relevant
to physics?  A degree of control comes precisely from the fact that
this class corresponds to regular Borel measures on $\Ab$. One can
thus rephrase the question of relevance of these measures in more
manageable terms: Is $\Ab$ ``too large'' to be physically useful and
mathematically interesting? We saw in this paper that, with this
rephrasing, the question can be analyzed quite effectively.  The
answer turned out to be in the negative; although $\Ab$ is very big,
it is small enough to admit a rich geometry. Finally, all our
geometrical structures naturally reside on the projective limit $\Ab$
of our family of compact analytic manifolds. It would have been
difficult and awkward to analyze them directly as generalized
structures on $\A$. Thus, there is no easy way out of making the
completions $\Ab$ and $\agb$.

To summarize, for diffeomorphism invariant theories, there is no easy
substitute for the extended spaces $\Ab$ and $\agb$; one has to learn
to deal directly with {\it quantum} configuration spaces. Fortunately,
this task is made manageable because there are three characterizations
of $\agb$: One as a Gel'fand spectrum of an Abelian $C^\star$-algebra
and two as projective limits. The three ways of constructing $\agb$
are complementary, and together they show that $\agb$ has a
surprisingly rich structure. In particular, differential geometry
developed in this paper makes it feasible to use $\agb$ to analyze
concrete mathematical problems of diffeomorphism invariant theories.

\bigskip\goodbreak
{\bf Acknowledgment}: We would like to thank John Baez, David
Groisser, Piotr Hajac, Witold Kondracki, Donald Marolf, Jose Mourao,
Tomasz Mrowka, Alan Rendall, Carlo Rovelli, Lee Smolin, Clifford
Taubes, and Thomas Thiemann for discussions.  Jerzy Lewandowski is
grateful to Center for Gravitational Physics at Penn State and Erwin
Schr\"odinger Institute for Mathematical Physics in Vienna, where most
of this work was completed, for warm hospitality.  This work was
supported in part by the NSF grants 93-96246 and PHY91-07007, the
Eberly Research Fund of Penn State University, the Isaac Newton
Institute, the Erwin Shr\"odinger Institute and by the KBN grant
2-P302 11207.

\bigskip
\goodbreak


\begin{thebibliography}{999}

\bibitem{AA1} A. Ashtekar, ``New variables for classical and 
quantum gravity'', Phys. Rev. Lett. {\bf 57}, 2244-2247 (1986);
``New Hamiltonian formulation of general relativity'', 
Phys. Rev. {\bf D36}, 1587-1603 (1987).

\bibitem{AM} M. Asorey and P.K. Mitter, ``Regularized, continuum 
Yang-Mills process and Feynman-Kac functional integral,'', Commun.
Math. Phys. {\bf 80}, 43-58 (1981); M. Asorey and F. Falceto,
``Geometric regularization of gauge theories'', Nucl Phys. {\bf B327},
427-460 (1989).

\bibitem{AI} A. Ashtekar and C.J. Isham, ``Representations of the 
holonomy algebras for gravity and non-Abelian gauge theories'', 
Class. \& Quan. Grav. {\bf 9}, 1433-1467 (1992).

\bibitem{AL1} A. Ashtekar and J. Lewandowski, ``Representation 
theory of analytic holonomy $C^\star$ algebras'', in {\it Knots and
quantum gravity}, J. Baez (ed), (Oxford University Press, Oxford
1994).

\bibitem{B1} J. Baez, ``Generalized Measures in gauge theory'', 
Lett. Math. Phys. {\bf 31}, 213-223 (1994).

\bibitem{B2} J. Baez, ``Diffeomorphism invariant generalized
measures on the space of connections modulo gauge transformations",
hep-th/9305045, to appear in the Proceedings of the conference on
quantum topology, D. Yetter (ed) (World Scientific, Singapore, 1994).

\bibitem{MM} D. Marolf and J. M. Mour\~ao, ``On the support of the 
Ashtekar-Lewandowski measure'', Commun. Math. Phys. (in press).

\bibitem{AL2} A. Ashtekar and J. Lewandowski, ``Projective techniques 
and functional integration for gauge theories'', preprint
CGPG-94/10-6, for the special JMP issue on functional integration,
edited by C. DeWitt-Morette.

\bibitem{KK} S. Klimek, W. Kondracki, ``A construction of two dimensional 
quantum chromodynamics'', Commun. Math. Phys.{\bf 113}, 389-402
(1987).

\bibitem{AA3} A. Ashtekar, {\it Non-Perturbative Canonical Gravity},
Lectures notes prepared in collaboration with R.S. Tate (World
Scientific, Singapore, 1991)

\bibitem{RS} C. Rovelli and L Smolin, ``Loop representation of quantum
general relativity'', Nucl. Phys. {\bf B331}, 80-152 (1990);
``Discreteness of Area and Volume in Quantum Gravity, gr-qc/9411005

\bibitem{ALMMT2} A. Ashtekar, J. Lewandowski, D. Marolf, J. Mour\~ao
and T. Thiemann, ``Quantization of theories of connections with local
degrees of freedom'', J. Math. Phys. (in press) 

\bibitem{AA2} A. Ashtekar, ``Recent mathematical developments in
quantum general relativity'', to appear in {\it The Proceedings of the
VIIth Marcel Grossmann Conference} R. Ruffini and M. Keiser (eds)
(World Scientific, Singapore, 1995). gr-qc/9411055

\bibitem{Le} J. Lewandowski, ``Group of loops, holonomy maps, path
bundle and path connection'', Class. Quant. Grav. {\bf 10},
879-904 (1993).

\bibitem{AMM} A. Ashtekar, D. Marolf and J. Mour\~ao, ``Integration on
the space of connections modulo gauge transformations'', in {\it The
Proceedings of the Lanczos International Centenary Conference}, J. D.
Brown, et al (eds) (SIAM, Philadelphia, 1994). 

\bibitem{K} A. N. Kolmogorov, {\it Foundations of the theory of 
probability} (Chelsea Publishing Company, N.Y. 1936).

\bibitem{DM} Y. Choquet-Bruhat, C. DeWitt-Morette and M. Dillard-Bleick,
{\it Analysis, Manifolds and Physics} (North Holland, Amsterdam, 1982);
section VIID. 

\bibitem{ALMMT1} A. Ashtekar, J. Lewandowski, D. Marolf, J. Mour\~ao
and T. Thiemann, ``A manifestly gauge invariant approach to quantum
theories of gauge fields'', in {\it Geometry of constrained dynamical
systems}, J. Charap (ed) (Cambridge University Press, Cambridge,
1994); ``Constructive quantum gauge field theory in two space-time
dimensions'' (preprint).

\bibitem{S} E. M. Stein, {\it Topics in analysis related to the 
Littlewood-Paley theory} (Princeton University Press, Princeton, 1970)
page 38.

\bibitem{ALMMT3} A. Ashtekar, J. Lewandowski, D. Marolf, J. Mour\~ao
and T. Thiemann, ``Coherent State Transforms for Spaces of Connections''
J. Funct. Analysis (in press) gr-qc/9412014.

\bibitem{Le2} J. Lewandowski, ``Topological measure and 
graph-differential geometry on the quotient space of connections'
Int. J. Mod. Phys. {\bf D3}, 207-210 (1994).

\bibitem{AL4}  A. Ashtekar, J. Lewandowski, in preparation.

\bibitem{Le11} J. Lewandowski, ``The loop quantization of gravity
in turms of the projective limit techniques'', in preparation.
\end{thebibliography}
\end{document}